\documentclass[fleqn,usenatbib]{mnras}

\usepackage{newtxtext,newtxmath}
\usepackage{comment}
\usepackage{graphicx}
\usepackage{graphics}
\usepackage{hyperref}
\usepackage{amsmath}
\usepackage{array}
\usepackage[utf8]{inputenc}
\usepackage{subcaption}

\usepackage[T1]{fontenc}

\graphicspath{{./}{Plots/}}

\DeclareRobustCommand{\VAN}[3]{#2}
\let\VANthebibliography\thebibliography
\def\thebibliography{\DeclareRobustCommand{\VAN}[3]{##3}\VANthebibliography}

\usepackage{graphicx}	
\usepackage{amsmath}	
\usepackage{layouts}
\usepackage{hyperref}

\newcommand{\astronomaly}{\textsc{astronomaly}}
\newcommand{\decals}{DECaLS}
\newcommand{\gz}{Galaxy Zoo DECaLS}

\usepackage{color}
\definecolor{purp}{RGB}{154,0,255}
\newcommand{\new}[1]{\textcolor{black}{#1}}
\newcommand{\km}[1]{\textcolor{black}{#1}}

\title[Enabling Unsupervised Discovery in Astronomical Images]{Enabling Unsupervised Discovery in Astronomical Images through Self-Supervised Representations}

\author[K. Mohale \& M. Lochner]{
Koketso Mohale,$^{1}$\thanks{E-mail: koketso.kjay@gmail.com}
Michelle Lochner,$^{1,2}$
\\
$^{1}$Department of Physics and Astronomy, University of the Western Cape, Bellville, Cape Town, 7535, South Africa\\
$^{2}$South African Radio Astronomy Observatory, 2 Fir Street, Black River Park, Observatory, 7925, South Africa\\
}

\date{Accepted 2024 March 26. Received 2024 March 19; in original form 2023 November 30}

\pubyear{2023}

\begin{document}
\label{firstpage}
\pagerange{\pageref{firstpage}--\pageref{lastpage}}
\maketitle

\begin{abstract}
Unsupervised learning, a branch of machine learning that can operate on unlabelled data, has proven to be a powerful tool for data exploration and discovery in astronomy. As large surveys and new telescopes drive a rapid increase in data size and richness, these techniques offer the promise of discovering new classes of objects and of efficient sorting of data into similar types. 
However, unsupervised learning techniques generally require feature extraction to derive simple but informative representations of images.
In this paper, we explore the use of self-supervised deep learning as a method of automated representation learning. We apply the algorithm Bootstrap Your Own Latent (BYOL) to \gz{} images to obtain a lower dimensional representation of each galaxy\new{, known as features}.
We briefly validate these features using a small supervised classification problem. We then move on to apply an automated clustering algorithm, demonstrating that this fully unsupervised approach is able to successfully group together galaxies with similar morphology. The same features prove useful for anomaly detection, where we use the framework \astronomaly{} to search for merger candidates. 
\new{While the focus of this work is on optical images, we also explore} the versatility of this technique by applying the exact same approach to a small radio galaxy dataset. This work aims to demonstrate that applying deep representation learning is key to unlocking the potential of unsupervised discovery in future datasets from telescopes such as the Vera C. Rubin Observatory and the Square Kilometre Array.
\end{abstract}




\section{Introduction}
Rapid advancements in astronomical surveys have resulted in the production of volumes of data too large for human experts to manually inspect. Projects such as Galaxy Zoo \citep{Lintott_2008} employ the use of citizen scientists to label galaxy morphology at large scales. Examples of applications of these morphology classifications include studies of the host galaxies of active galactic nuclei \citep{gz_for_agn}, investigations of merging galaxies \citep{gz_for_merger}, and the formation of bars in spiral galaxies \citep{gz_for_bars}. However, these classifications are limited in that they require hand-labelled data and the rapid increase of data volumes means that eventually even citizen science projects will not be able to keep up. Additionally, citizen scientists may not necessarily have the training required to make more nuanced classifications or identify particularly anomalous sources.

Unsupervised machine learning has the potential to leverage large unlabelled datasets, allowing for data-driven discovery and exploration. These algorithms can also be used to create training sets for downstream supervised applications far more quickly than random labelling. There have been numerous applications of unsupervised learning in astronomy, including the application of clustering and anomaly detection to radio galaxies \citep{Ralph2019, Gupta2022}, the detection of anomalous spectra in LAMOST data \citep{Yang2023} and the hunt for rare transients and variables in Deeper Wider Faster optical data \citep{webb2020}. \citet{Lochner2021} introduced the general-purpose anomaly detection framework \astronomaly{}, which was later applied to data from the MeerKAT telescope to discover a highly unusual radio source \citep{lochner2023unique}.

There is also a long history of applying unsupervised techniques to achieve automatic clustering of optical galaxies into similar morphology, which include early applications of artificial neural networks to photometric galaxy parameters \citep[e.g.][]{Lahav1996, Naim1997, dAbrusco2007}. The motivation behind these efforts is not simply to automate otherwise tedious tasks, but also to allow the possibility of discovering new classes of objects or a more physically motivated categorisation of galaxies. 

However, unsupervised algorithms usually cannot work directly with high dimensional data such as images \citep[with the notable exception of the rotation-invariant self-organising maps of][]{polsterer2019} and instead require lower dimensional representations called features. In the case of galaxy morphologies, these features could include photometric colours or shape parameters such as the Gini coefficient. The choice of feature extraction method will generally dictate the success of the machine learning algorithm. 

Deep learning algorithms \citep[e.g.][]{LeCun2015}, particularly convolutional neural networks (CNNs), have revolutionised the field of image recognition due to their ability to obtain meaningful representations from images without requiring explicit feature selection. This has led to their use as general-purpose feature extractors, as was demonstrated in \citet{Walmsley_2022} where a pretrained convolutional neural network dramatically outperformed simpler, hand-designed morphological features for unsupervised tasks on the optical dataset \gz{} \citep{galaxyzoo}. \citet{etsebeth2023} recently demonstrated the same features could be used for highly effective anomaly detection among nearly 4 million galaxies in the larger \decals{} dataset \citep{Dey_2019}.

The downside of using a pretrained network as a feature extractor is that it requires the data be relatively similar to the dataset it was originally trained on in order to obtain the best performance. It also requires a large, labelled dataset for training which does not exist for many fields, such as high resolution radio astronomy. In a recent paper, \citet{sadr2022} applied a CNN to the \gz{} dataset, repeatedly retraining the network with human-provided labels to improve the features learned in order to quickly detect interesting anomalies. While this approach is very promising, it would require somewhat expensive retraining of the CNN with each iteration which may not be possible in all scenarios.  

Self-supervised learning offers a promising alternative. Modern approaches to unsupervised clustering of galaxy images tend to rely on autoencoders to learn a representative feature space \citep[for example,][]{Spindler2021, Cheng2021, Zhou2022, Fielding2022}. Autoencoders are trained to reproduce the input data identically and in the process, learn a lower dimensional representation of the image dataset. Newer self-supervised techniques, including contrastive and non-constrative learning, instead apply augmentation to introduce random variations in the data and train the algorithm to recognise these pairs of augmented images as being the same. This form of self-supervised learning is gaining popularity as a method of learning effective representations without labels \citep{contrastivelearning}. 

Contrastive learning, which includes both positive and negative augmented image pairs, has been applied several times to optical galaxy images. \citet{Hayat_2021} and \citet{Stein2021, Stein2022} use contrastive learning for an impressive array of applications including galaxy morphology classification, photometric redshift estimation, strong lens discovery and similarity searches. \citet{Sarmiento2021} applied contrastive learning to investigate galaxy physics in integral field unit data, while \citet{Wei2022} showed that learned representations can be applied across different optical datasets.

While this recent body of work shows that contrastive learning can be a powerful tool for a myriad of applications, it can be computationally demanding. Non-contrastive learning, specifically the technique Bootstrap Your Own Latent \citep[BYOL,][]{grill2020bootstrap}, is a more viable option when computational resources are limited. \citet{slijepcevic2023radio} showed the potential of BYOL as a foundation model for radio galaxy images, which can be fine-tuned for classification tasks on new datasets.

The goal of this paper is to develop a methodology for data exploration, as an initial step for building training sets, obtaining the more obvious morphology groups and for detecting anomalies for large and unlabelled datasets. We apply self-supervised learning to two different datasets, for the purpose of learning representations that can be used for unsupervised learning tasks. As well as demonstrating the performance of these representations for a simple supervised classification task, we apply automated unsupervised clustering and anomaly detection with active learning as example downstream tasks. While the main dataset used in this paper consists of optical images of galaxies, we also apply the same methods to a small radio galaxy dataset to demonstrate their general utility. 

We start by focusing on the popular hand-labelled optical dataset \gz{}, which we describe in \autoref{sec:galaxy_data}. We outline our use of a pretrained network for removing artefacts in \autoref{sec:pretrained} and the methodology of using self-supervised learning to extract useful features  in \autoref{sec:methodology}. We then demonstrate the effectiveness of these features in a series of different applications. We make use of supervised classification in \autoref{sec:evaluation} to demonstrate the utility of initialising the network with pretrained weights and as a baseline to ensure the self-supervised method is indeed learning representations that correspond to galaxy morphology. An unsupervised clustering approach is explored in \autoref{sec:clustering} to attempt to group galaxies of similar morphology together without the need for labelling. We apply an anomaly detection algorithm in \autoref{sec:anomaly_detection} to rapidly locate merger candidates. After exploring the use of self-supervised learning on optical data, we further demonstrate its remarkable flexibility by applying an essentially identical approach to a radio dataset in \autoref{sec:mirabest} with both a supervised classification and clustering application. \autoref{sec:conclusions} summarises our conclusions. 

\section{Galaxy Zoo DECaLS Data}
\label{sec:galaxy_data}
The data used in this paper was originally sourced from the Dark Energy Camera Legacy Survey \citep[\decals{},][]{Dey_2019} DR5. The \gz{} data \citep{galaxyzoo} is a catalogue of thousands of high resolution images, from \decals{}, of optical galaxies with a wide range of morphologies. This dataset is ideal for our work because it is large, from a modern telescope and fully labelled. Although we are primarily interested in unsupervised applications, the labels allow us to test the effectiveness of the algorithms considered before applying to unlabelled datasets in the future. The Galaxy Zoo\footnote{\url{www.galaxyzoo.org}} citizen science project  asks users to identify morphological features of galaxies, deciding whether each object is smooth or featured, has spiral arms, bars, tidal tails etc. \gz{} made use of a sophisticated decision tree, rather than assigning simple morphological labels, which simplifies the identification task by removing jargon but also allows fine-grained decision making when defining a morphological sample. 

Volunteers are presented with a series of questions such as \emph{``Is the
galaxy simply smooth and rounded, with no sign of a disk?''} to which the answer could be ``Smooth'', ``Features or disk'' or ``Artefact''. The volunteer is then presented with the next question based on their initial answer and so traverses down the decision tree. The number of votes as well as the vote fraction for each question is recorded and these can be used to create labelled subsets that select for a certain morphology.

The images and labels we use can be found on Zenodo\footnote{\url{https://zenodo.org/record/4573248}} \citep{walmsley2020_zenodo}. We use the labels from versions 1 and 2 of \gz{}\footnote{\href{https://zenodo.org/record/4573248/files/gz_decals_volunteers_1_and_2.parquet?download=1}{\texttt{gz\_decals\_volunteers\_1\_and\_2.parquet}}} rather than version 5, as we did not consider the improvements to the decision tree in version 5 to be as important for our application as simply having more labelled data to test with. The total number of images available from Zenodo was 269 760 and of those, 65 290 had labels from volunteers. We used the full image set for training our feature extractor (\autoref{sec:methodology}) and applying a clustering algorithm (\autoref{sec:clustering_methodology}) but restricted our analysis to the labelled data for interpretation of the clustering results (\autoref{sec:clustering_results}) and the later anomaly detection application (\autoref{sec:anomaly_detection}).

\subsection{Preprocessing}
\label{sec:preprocessing}
\citet{slijepcevic2023radio} found that CNNs are sensitive to the apparent angular size of a source in a given image. To ensure the algorithm is focusing on physical morphological features, we preprocess each image to attempt to isolate and resize the central source. We use the standard sigma clipping transform available in the software package \astronomaly{}\footnote{\url{https://github.com/MichelleLochner/astronomaly}} to locate the central source and cut out background sources. This function works by first calculating the noise level in the image, using \textsc{Astropy} \citep{Astropy1,Price-Whelan_2018,TheAstropyCollaboration_2022}, applying a $4\sigma$ threshold and then using an \textsc{openCV} \citep{bradski2000} contour-finding algorithm to select all regions above this threshold. Because the source should always be located at the centre of the image, we could then select only the central contour in order to determine an appropriate bounding box around the source. We chose to enlarge the bounding box by a factor of two to ensure the entire source is contained. 

It should be noted that the resulting bounding box was applied to the original image, not the sigma-clipped image. Sigma clipping can sometimes remove part of the source and a key advantage of deep learning is that it can easily learn to ignore the background anyway. \new{We thus elected not to use sigma clipping for this analysis, but if this methodology were to be applied to more crowded fields it may become necessary to refine the preprocessing to remove nearby sources}.

At times the contour-fitting procedure can fail and raise an error, usually due to large sources filling the field or very bright nearby sources such as stars or artefacts. In these cases we simply use the original image instead. After extracting the central source, we resize the image to 300x300 pixels. 

\section{Feature extraction with a pretrained network}
\label{sec:pretrained}
Our general approach is to make use of a CNN as a feature\footnote{\new{It should be noted that these features are representations of the images derived from a neural network and do not refer to the more general use of the word feature in the context of galaxy morphology such as ``bar'' or ``tidal tail''.}} extractor, rather than a classifier. This can be done for a CNN that has been trained to solve a different task by ignoring the final classification layer of the network and instead using the outputs of the weights of the previous layer as features (as was done in \citet{Walmsley_2022}, \citet{etsebeth2023} and several other examples). Here we describe the core architecture we use throughout this work and our initial experiment with a standard pretrained neural network before moving on to self-supervised learning.

\subsection{Model architecture}
CNNs are a type of neural network consisting of many layers of interconnected ``neurons''. \new{Standard artificial neural networks take numerical input and, through a series of non-linear operations performed by the neurons, make predictions such as the class the input data belongs to.} The key difference between CNNs and typical artificial neural networks is that the neurons of CNNs are kernels that perform convolutions across an image. This allows the network to, through training, learn an optimal set of filters through which to pass the image, resulting in a useful image representation for the downstream task. CNNs also consist of other types of layers, such as pooling and dropout layers, which are inserted to improve this representation by allowing filters to be applied on a hierarchy of scales and also avoiding overfitting, which such complex algorithms are otherwise prone to. \new{CNNs are typically used for classification tasks, but the majority of the network architecture is actually dedicated to learning features that are ultimately useful for classification. We can leverage this by using a pretrained CNN as a general-purpose feature extractor.}

\begin{figure*}
    \centering
    \includegraphics[width = \linewidth]{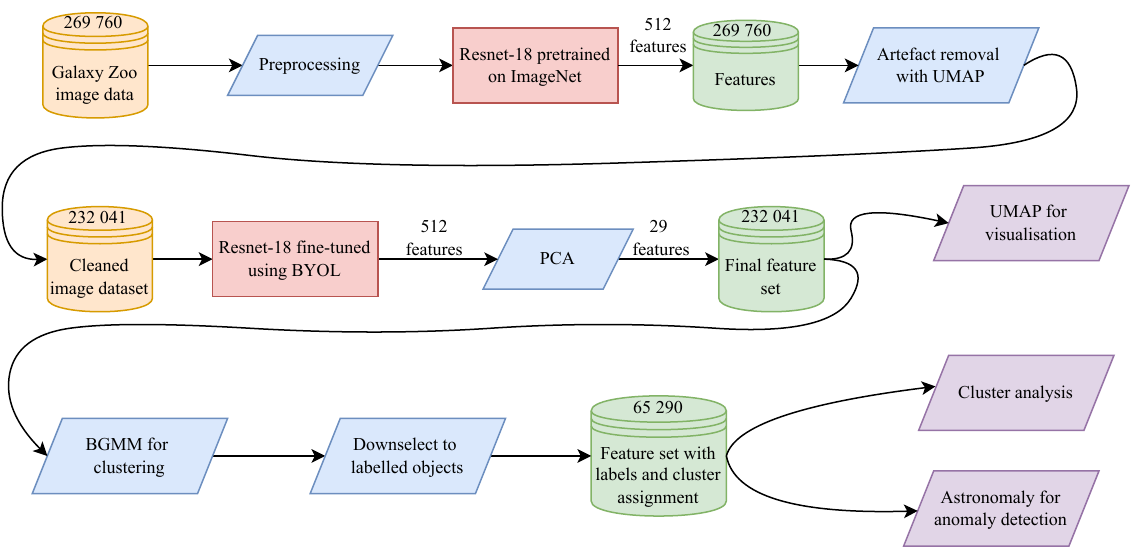}
    \caption{\new{Flow diagram of our methodology. Data and features are represented as cylinders, processes as parallelograms and neural networks as rectangles. The numbers on top of the cylinder symbol indicate how many objects are in that dataset, whether as images or rows in the feature array. Where dimensionality is reduced, the resulting number of features is also indicated on the arrows.}}
    \label{fig:flow_diagram}
\end{figure*}

We used the ResNet-18 model \citep{resnet} as the CNN architecture for this work. ResNet-18 was found to have sufficient performance for a low computational cost. \citet{second_last_layer} performed an in-depth analysis of the use of CNNs as feature extractors for clustering tasks and found that the second last layer always provides the best representation of the images. In the case of a Resnet-18 this is the layer called ``avgpool'', which produces 512 outputs to be used as features. \new{\autoref{fig:flow_diagram} illustrates our full methodology, showing how features are extracted from the raw data.}

\subsection{Visualisation of extracted features}
\label{sec:umap}
It is particularly critical for this type of work to be able to visualise high dimensional feature spaces. In keeping with the current trend in the machine learning field, we make use of the technique Uniform Manifold Approximation and Projection \citep[UMAP,][]{umap}. UMAP aims to learn a lower dimensional embedding that optimises for local structure, but still preserves global structures. \new{The UMAP algorithm constructs a fuzzy topological representation of the data and then approximates this with a lower dimensional representation through an optimisation algorithm. This allows the structure of high dimensional data to be easily visualised.} These plots are especially useful for localising obvious outliers, determining where sources of particular types lie in feature space and for understanding the behaviour of the unsupervised algorithms we later applied to the features.

To implement UMAP for our features, we made use of the \textsc{umap-learn}\footnote{\url{https://umap-learn.readthedocs.io/en/latest/}} software package \citep{sainburg2021}. The UMAP algorithm has a number of hyperparameters. Throughout this work, we set the parameter ``number of neighbours'' to 15, which minimises the creation of artificial clusters and the parameter ``minimum distance'' to 0.01 to prioritise local structure and encourage the formation of genuine clusters. \new{We demonstrate the effect of different parameter choices in the Appendix, \autoref{appendix:umap}.}

\subsection{Artefact removal with a pretrained network}
\label{sec:imagenet}
\begin{figure}
    \centering
    \includegraphics[width = \linewidth, trim={1cm 1cm 1cm 1cm}, clip]{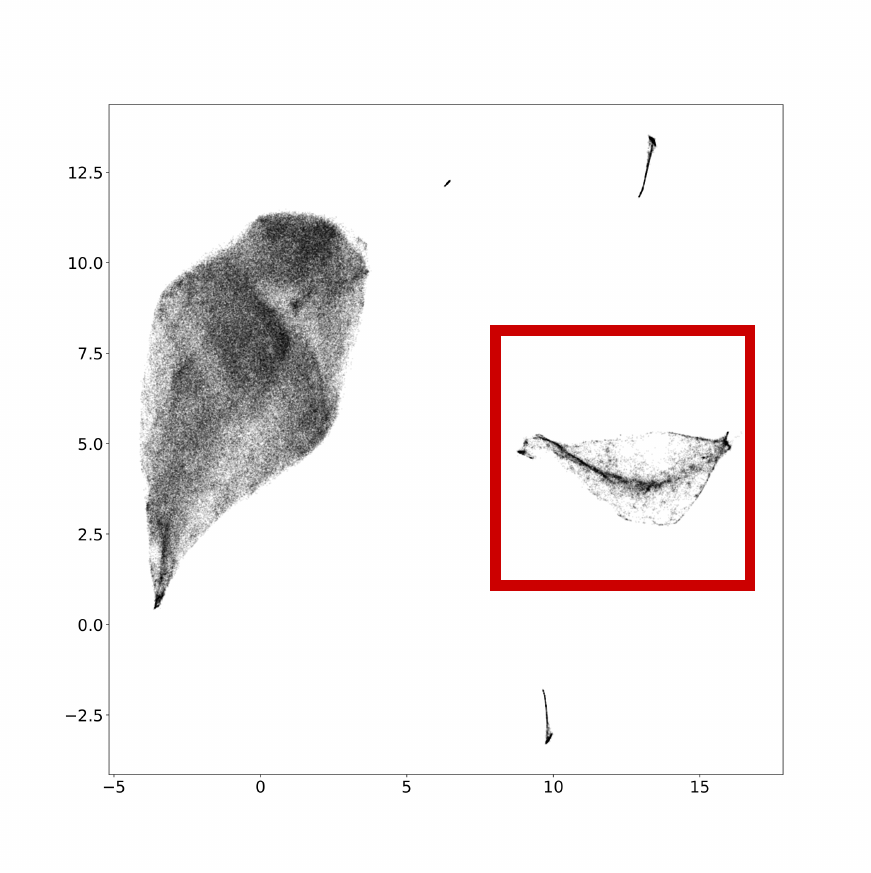}
    \caption{UMAP representation of the feature space of the whole \gz{} data set after preprocessing, using a CNN pretrained on ImageNet as a feature extractor. The bounding box shows the selection used to remove the artefacts introduced. The $x$ and $y$ axes are in arbitrary units.}
    \label{fig:umap_imagenet}
\end{figure}

Unfortunately, the preprocessing procedure of \autoref{sec:preprocessing} inadvertently introduced artefacts into the data \new{(see \autoref{appendix:artefacts} in the Appendix for more details of how the artefacts are created)}. While a more refined procedure may produce fewer artefacts, it is difficult to eliminate them entirely. Instead, we used the opportunity to test a novel approach to artefact removal. Before proceeding to the full feature extraction method described in the next section, we used a ResNet-18 already trained on the well-known machine learning dataset of terrestrial images called ImageNet \citep{imagenet}. This pretrained model is already a fairly effective feature extractor and we were able to trivially excise the most obvious artefacts from our dataset with these features. A UMAP plot is shown in \autoref{fig:umap_imagenet} displaying the cuts applied to excise the artefacts. We focused on the largest cluster of artefacts rather than trying to vigorously remove all artefacts. While the pretrained network was highly effective at removing these, it failed to usefully group similar galaxies together. We thus did not make use of this network, beyond the initial artefact removal, and instead moved on to a self-supervised learning technique as a feature extractor.
The main dataset used in the rest of this work contains 232 041 images after preprocessing. \new{The artefact removal process is illustrated in the flow diagram of \autoref{fig:flow_diagram}.}

\section{Feature extraction with self-supervised learning}
\label{sec:methodology} 
Self-supervised learning aims to learn useful representations of images without requiring training labels. A common approach to self-supervised learning is to train the model to have the same predictions for different augmentations (views) of the same image. This leads to representation collapse (e.g the model might predict the same trivial solution for all images) and self-supervised learning techniques employ different ways to circumvent this. For example, the contrastive learning algorithm SimCLR \citep{simclr} uses a repulsive term, generated from negative pairs, in the loss function to prevent collapse. However, this algorithm can be resource-intensive precluding its use for this work.

\begin{figure}
    \centering
    \includegraphics[width = \linewidth]{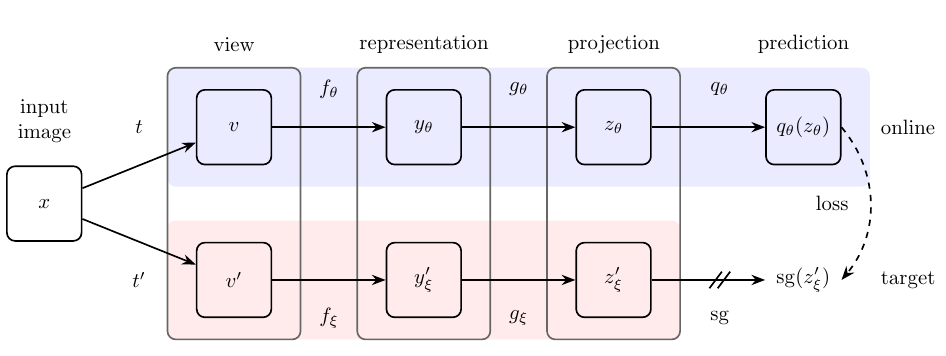}
    \caption{The Bootstrap Your Own Latent architecture \citep{grill2020bootstrap}.}
    \label{fig:byol}
\end{figure}

Bootstrap Your Own Latent \citep[BYOL,][]{grill2020bootstrap}, on the other hand, is a non-contrastive self-supervised learning method that requires relatively low computational resources and, somewhat surprisingly, manages to avoid representation collapse without the need for negative pairs (see \citet{Tian2021} for some recent insights and possible explanations). BYOL uses two neural networks of the core same architecture, called the online and target networks, \new{shown in \autoref{fig:byol}}. \new{The goal of the online network is, given an augmented view of an image, to predict the output of the target network, which is given a \emph{different} augmented view of the same image. The representation indicated in \autoref{fig:byol} is the final layer of a standard CNN (such as Resnet-18), similar to what was used in \autoref{sec:imagenet} to remove artefacts. The representations of the online network is eventually what is used as features, the rest of the architecture being discarded. The projection layer is a simple fully connected neural network layer to reduce the dimensionality of the representations. This projection layer gives the online network something to predict that is more tractable than the representation layer. The online network updates its weights through a standard gradient descent algorithm, minimising the mean squared error between the two projection layers.} The weights of target network ($\xi$) are an exponential moving average of the online network weights ($\theta$) and are updated according to equation \autoref{eq:target_params}. Here $\tau \in [0,1]$ is a target decay parameter.

\begin{equation} 
\xi_i  \longleftarrow \tau \xi_{i-1} + (1- \tau)\theta
\label{eq:target_params}
\end{equation} 

\new{Through this update step, the two networks should converge to similar representations for two different augmented views of the same image.} By training on a large dataset with random augmentations, this pair of interacting networks is able to produce a representation of the images that reliably groups similar objects together without the need for labels.

\subsection{Augmentations and hyperparameters}
\label{sec:augmentations}
We use the package \textsc{byol-pytorch}\footnote{\url{https://github.com/lucidrains/byol-pytorch}} \citep{chen2020} to apply the BYOL algorithm. The hyperparameters we selected are shown in \autoref{tab:byol_hyperparams}. 
We fix the number of epochs to 20, which we found to be computationally efficient while still being sufficient to produce excellent representations. Further training did not improve our results. \km{The batch size was set to 128 because it was the largest possible given our computational resources and the learning rate to 0.0001, which was found to be optimal for this dataset (see the Appendix, \autoref{appendix:byol_hyperparameters}). We follow the default settings from \cite{grill2020bootstrap} for all other hyperparameter settings.} We chose not to use a separate validation set to maximise the number of sources available for training, instead relying on the downstream supervised learning application to assess the features (see \autoref{sec:evaluation}).

 The effectiveness of BYOL is heavily dependant on the augmentations. \citet{grill2020bootstrap} includes recommended augmentations to use for terrestrial dataset. We find that, just as in the case of radio data \citep{slijepcevic2023radio}, augmentations have to be adjusted to be suitable for the optical datasets. Preprocessing and augmentations affect the representations that BYOL will learn and it is important that the variance in the background of galaxies does not dominate the variances in different morphology types. 

 \begin{table}
    \centering
    \begin{tabular}{cc}
    \hline
    Hyperparameter & Value \\
    \hline
    Network architecture & Resnet-18 \\
    Epochs & 20 \\
    Optimiser & Adam \\
    Learning rate & 0.0001 \\
    Batch size & 128 \\
    $\tau$ & 0.99 \\
    Neurons in projection layer & 256 \\
    
    \hline
    \end{tabular}
    \caption{Hyperparameter values used for training the BYOL algorithm.}
    \label{tab:byol_hyperparams}
\end{table}

\new{Using the analysis performed in \citet{slijepcevic2023radio} as a starting point, we selected the augmentations gaussian blurring, vertical flip, horizontal flip, each with probability 0.5. We also applied the augmentation resized crop with the smallest value for cropping the image set to 0.7, as well as rotations by random angles $\theta \in [0-360]$, setting the probability of applying either to 0.7. While a full ablation study would be too computationally intensive, we were able to investigate the importance of each augmentation in the Appendix, \autoref{appendix:byol_hyperparameters}, by applying only one augmentation at a time and examining the accuracy on a downstream supervised learning task. We find the highest performance by applying all the augmentations listed. Similar to what was found in \citet{grill2020bootstrap} and \citet{slijepcevic2023radio}, the exact choice of augmentations has only a moderate impact on KNN accuracy. However it can have a larger impact on the versatility of the final representation and visual inspection showed that using all augmentations produced features that better grouped similar sources together than each augmentation alone.}

In terms of computational requirements, we use a single  NVIDIA P100, 16-core GPU (with 116GB of RAM) and find that BYOL trains in approximately 11 hours. 

\subsection{Dimensionality reduction}
Most unsupervised learning algorithms do not scale well to high dimensional spaces. For this reason, we further reduced the dimensionality of the features using Principal Component Analysis \citep[PCA,][]{Pearson1901, Hotelling1933}.  \new{PCA works by decomposing a dataset into a new coordinate space such that each orthogonal component vector aligns with directions of progressively decreasing variance. By keeping only a small number of principal components, the majority of information can be retained with a dramatic reduction in dimensionality. PCA is especially valuable for highly correlated variables, such as the features obtained from training a neural network.}

It is natural to consider using the reduced features derived from a manifold learning algorithm such as t-SNE \citep{tsne} or UMAP \citep{umap} instead of PCA, since these algorithms are non-linear and more flexible. However, we found that PCA preserves \emph{global} structure better than manifold learning approaches and it is precisely the linearity of PCA that reduces the risk of creating artificial clusters in feature space (which manifold learning can sometimes produce). We thus elected to use PCA to reduce dimensionality for downstream tasks and UMAP purely for visualisation purposes. 

We applied PCA to the deep representations, as obtained from the hidden layer ``avgpool'', keeping 95\%\ of the variance and thus reducing to 29 principal components. We exclusively used the reduced feature space for clustering and anomaly detection. \new{The second row of the flow diagram in \autoref{fig:flow_diagram} shows how features are extracted from the main dataset.}

\section{Evaluation of Extracted Features}
\label{sec:evaluation}
\subsection{Evaluation subset}
\label{sec:data_labels}
It is challenging to evaluate the performance of a self-supervised learning algorithm since they are designed to operate without any labels by definition. The standard approach is usually to evaluate the extracted features in a downstream task. We thus opted to use our features to solve a simple, if contrived, supervised classification problem as an initial test of performance. 

We selected a small sample of sources that should be considered relatively easy to classify: round ellipticals, spiral galaxies and edge-on galaxies. Because Galaxy Zoo makes use of a decision tree rather than hard morphological classifications, cuts must be used to extract a confident sample of sources, which we describe below. In every case, we ensured a minimum number of five votes for the question being considered (the same threshold used in \citet{Dominguez2018}). We also selected only galaxies most likely not to host a merger, by requiring \texttt{merging\_merger\_fraction} $< 0.2$. The number of each class that meets the cuts is given in brackets. 

\begin{figure}
    \centering
    \includegraphics[width = \linewidth]{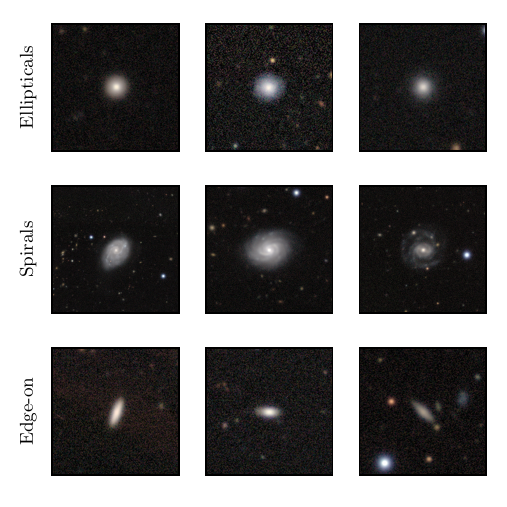}
    \caption{Examples of each class in the evaluation set representing ellipticals (top row), spirals (middle row) and edge-on galaxies (bottom row).}
    \label{fig:examples}
\end{figure}

\begin{itemize}
\item Round ellipticals (4231): \\ \texttt{smooth-or-featured\_smooth\_fraction} $>0.8$ and \\ \texttt{how-rounded\_completely\_fraction} $>0.8$.
\item Spirals (4034): \\ \texttt{smooth-or-featured\_featured-or-disk\_fraction} $>0.8$ and \\ \texttt{has-spiral-arms\_yes\_fraction} $>0.8$.
\item Edge-on galaxies (5344): \\ \texttt{disk-edge-on\_yes\_fraction} $>0.8$.
\end{itemize}
\autoref{fig:examples} shows three randomly chosen examples for each of the classes.

To evaluate the utility of our extracted features for this three-class classification problem, we applied a simple k-nearest neighbours (KNN) algorithm \citep{Fix1951, Cover1967}. For this problem, we found that KNN performed just as well on the original feature space as on the PCA-reduced space. Thus, all reported results using KNN are applied in the original 512-dimensional space. 

The \textsc{scikit-learn} \citep{scikit-learn} implementation of KNN was used and we set the number of neighbours to 5 and the distance metric to Minkowski. We selected the overall accuracy as an easily-interpretable metric, given that the classes are approximately balanced. As our focus is not on supervised learning and we simply use this as a tool to evaluate our features, we did not further optimise the classifier. In all cases, the KNN algorithm was trained on a randomly chosen 75\% of the evaluation subset and the accuracy was computed on the remaining 25\%. For each epoch we run KNN 50 times with a new random training-test split for each iteration and investigate the mean and standard deviation to ensure the algorithm converges to a roughly constant accuracy. 

While the accuracy is a useful tool to monitor the performance of the BYOL algorithm as a function of epoch, it's important to note that the algorithm is still trained in a completely self-supervised way and that this accuracy information is never fed back to the network. 

\subsection{Comparing transfer learning and random weight initialisation}
\label{sec:supervised}
It is most common to train a self-supervised learning algorithm with the weights initialised to random values. However, inspired by the performance of the pretrained network in \citet{Walmsley_2022} and many other examples of fine-tuning, we decided to compare the performance of BYOL initialised with random weights with that of ImageNet-intialised weights. We thus essentially use BYOL as a method of fine-tuning the network described in \autoref{sec:imagenet} to adapt the network to our particular dataset.

\begin{figure}
    \centering
    \includegraphics[width = \linewidth]{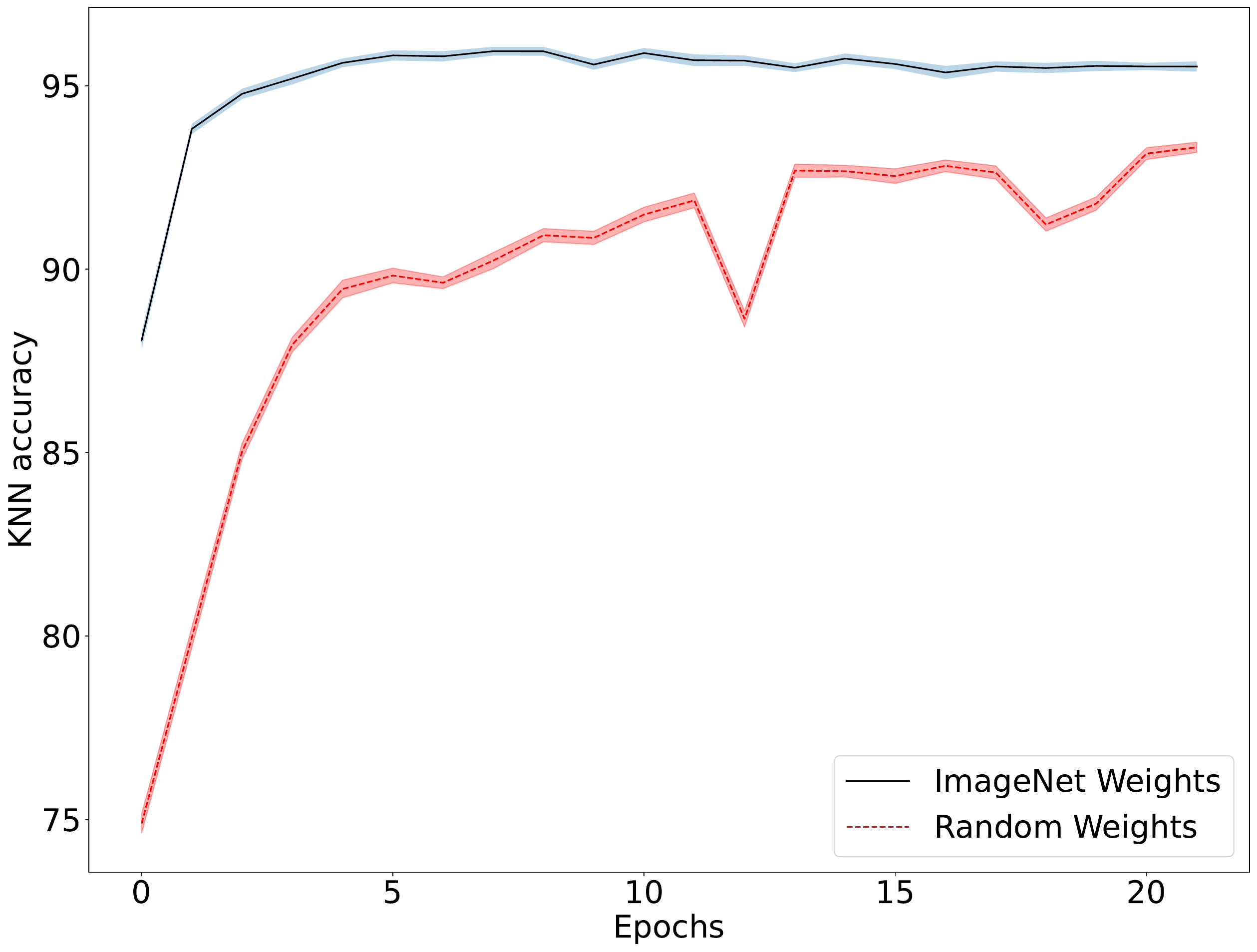}
    \caption{Accuracy of a KNN algorithm applied to the \gz{} evaluation subset as a function of epoch. We use the features derived from training BYOL on the galaxy images and compare initialising the weights randomly with initialising them from a network pretrained on ImageNet. For each epoch, the accuracy is computed for 50 iterations of training-test splits with the mean represented as a line and the standard deviation as an envelope. Fine-tuning increases the accuracy and convergence speed for this dataset. \new{The corresponding F1 scores (harmonic mean of the precision and recall) are listed in \autoref{tab:f1_galaxy} in the Appendix.}}
    \label{fig:knn_accuracy_for_galaxyzoo}
\end{figure}

\autoref{fig:knn_accuracy_for_galaxyzoo} shows the accuracy of KNN applied to the evaluation set when training a model initialised with ImageNet weights against one with randomised initial weights. The results show that not only do the fine-tuned BYOL features start with a 13\%\ higher accuracy, the network also converges quicker to a relevant feature space and has a higher peak accuracy after 20 epochs. While the accuracy of the random-start network may eventually be comparable to that of the fine-tuned network, fine-tuning provides significant gains in terms of reducing computational requirements. While we use the accuracy purely as a comparison metric between the two approaches, it is also worth noting that 95\%\ accuracy is excellent performance and shows the representation produced by BYOL is useful for downstream tasks. Having evaluated its performance, for the remainder of this work we use the features derived from applying BYOL, initialised with the ImageNet weights, to the entire \gz{} dataset after preprocessing.

\section{Clustering}
\label{sec:clustering}

After extracting features, evaluating their performance on a small supervised learning problem and reducing their dimensionality with PCA, we next turned our attention to the main goal of this paper: to automatically cluster similar objects together in an unsupervised manner. 

\subsection{Bayesian Gaussian mixture model}
\label{sec:clustering_methodology}
We chose to apply a Bayesian Gaussian mixture model \citep[BGMM,][]{Attias1999} for this problem. The aim of this clustering approach is to approximate the data as a mixture of Gaussian distributions, each with a mean, covariance and overall weight. These Gaussians then form the clusters. A BGMM specifically applies Bayesian inference to learn the parameters of these Gaussians. \new{Rather than computing the full posterior over the parameters, which is computationally expensive, the algorithm we selected uses variational inference to approximate the posterior distribution and find the best-fitting parameters through optimisation.}

 We use the \textsc{BayesianGaussianMixture} library from \textsc{scikit-learn} \citep{scikit-learn}. We ran BGMM with the number of components set to 20. The intuition behind the number is based on the number of density regions we observed on the UMAP feature space. However, we note that the BGMM implementation is able to set the weights of individual Gaussians very low making the input number of components an upper limit in reality. We set the weight concentration prior to 0.5 to force BGMM to focus more on global structure. \new{It is challenging to tune this hyperparameter in general in an unsupervised context. The simplest approach, which we employed, is to vary the weight concentration prior and visually inspect the resulting clusters. A poor choice of this parameter results in obvious overlap between clusters and mixing of source types within clusters.} \km{We set the number of initialisations to 10 to control the number of times the algorithm will runhttps://github.com/MichelleLochner/astronomaly to ensure reproducibility. We also increase the total number of iterations per run to 1000 to ensure that the algorithm converges in each run. All other hyperparameters we kept to their default}

\begin{figure*}
    \begin{minipage}{\linewidth}
    \centering
    \includegraphics[width = 0.9\linewidth]{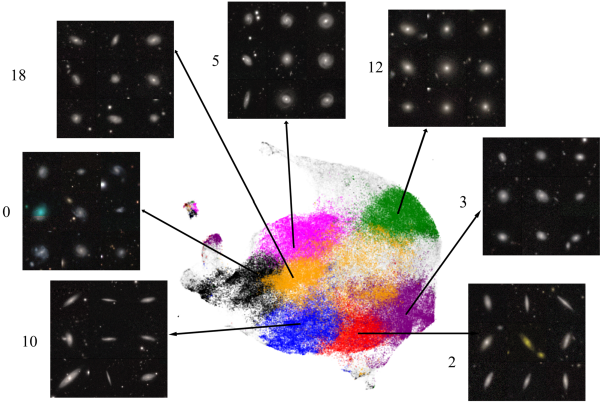}
    \caption{UMAP plot of cluster samples corresponding to morphologies that include flat disk galaxies (2 and 10), bright centered tight spirals (5 and 18), and bright centered high resolution ellipticals (12).}
    \label{fig:clean_cluster_samples}

    \end{minipage}

\end{figure*}

\begin{figure*}
    \begin{minipage}{\linewidth}
    \centering
    \includegraphics[width = 0.9\linewidth]{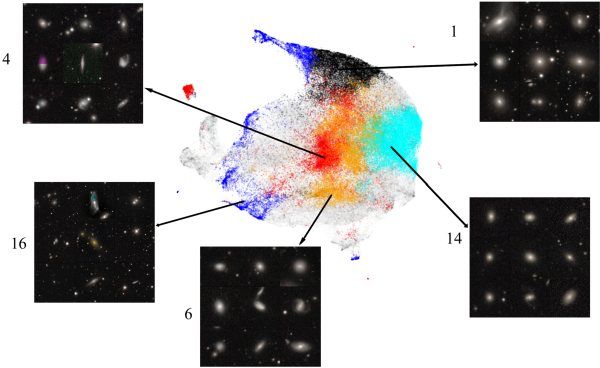}
    \caption{UMAP plot of cluster samples corresponding to morphologies that can be challenging to categorise. The feature space shows a larger disagreement between UMAP and BGMM in the case of these type of galaxies. The confusion mostly comes from heavy background noise (cluster 16, 4 and 1) as well as low resolution (cluster 14).}
    \label{fig:non_clean_cluster_samples}
    \end{minipage}
\end{figure*}

\begin{figure*}
    \begin{minipage}{\linewidth}
    \centering
    \includegraphics[width = \linewidth]{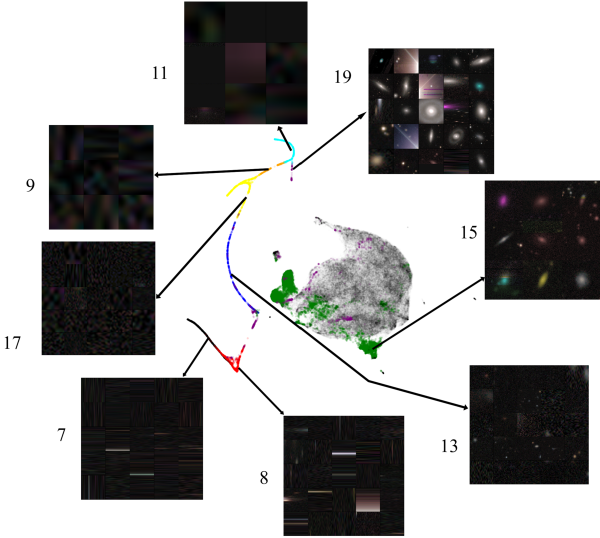}
    \caption{UMAP plot of cluster samples that correspond to artefacts. These artefacts tend to separate well from the main population and group together by type, indicating that self-supervised features are effective at identifying artefacts that may have been otherwise missed.}
    \label{fig:artefacts_cluster_samples}
    \end{minipage}
\end{figure*}
 
We applied the clustering algorithm to the features from the full \gz{} dataset (after preprocessing) to obtain a total of 20 clusters labelled 0-19. BGMM is able to identify densities consistent with those that we see on the feature space in the UMAP plots. However we observe some disagreement between BGMM and UMAP since the clustering is applied directly on the principal components. 

\autoref{fig:clean_cluster_samples} shows examples of galaxies that correspond to several of the clusters that can be seen on the UMAP plot. Combining self-supervised with unsupervised learning does appear to successfully group together sources with similar morphologies, such as elliptical and spiral galaxies, as well as edge-on galaxies and those with a prominent bulge. \autoref{fig:non_clean_cluster_samples} shows some more complex examples, suggesting that at times the algorithm may be picking up on non-physical properties such as a particularly ``zoomed out'' image or the presence of a companion. Improved preprocessing could perhaps clean up some of these clusters but blending will likely remain a difficult challenge to these algorithms as surveys increase in depth and hence source density.

\autoref{fig:artefacts_cluster_samples} shows how the clustering algorithm effectively selects artefacts, missed by the approach presented in \autoref{sec:imagenet}. This suggests that unsupervised techniques could provide a relatively lightweight method of detecting and removing artefacts that may be otherwise missed by automated pipelines.

\subsection{Cluster analysis with volunteer labels}
\label{sec:clustering_results}
While it is visually apparent that the clustering algorithm successfully groups galaxies of similar morphology together, we can investigate this more deeply by making use of the Galaxy Zoo decision tree user labels. \new{The bottom row of the flow diagram in \autoref{fig:flow_diagram} shows how only objects with labels are selected and how those subsequent features are used for clustering and later, anomaly detection (\autoref{sec:anomaly_detection}).}

\begin{figure*}
    \centering
    \includegraphics[width = 0.92\linewidth]{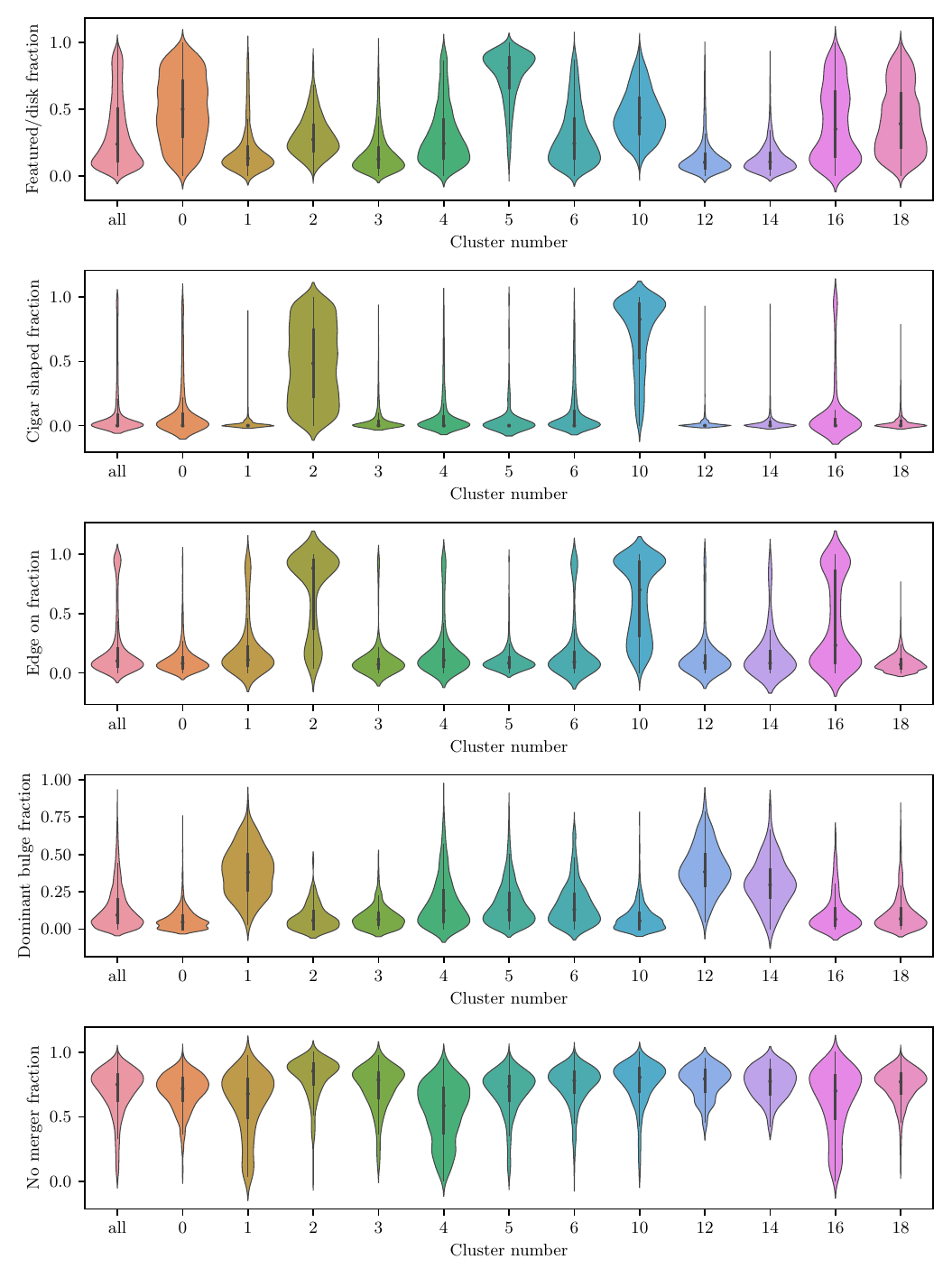}
    \caption{Vote fraction distributions for various Galaxy Zoo questions for each cluster (clusters of predominantly artefacts are not plotted). The global distribution is indicated with the ``all'' label. The distributions are represented as violin plots (using the package \textsc{Seaborn} \citep{Waskom2021}): a kernel density estimate of the distribution with lines representing the extrema, a box for the first and third quartiles and a point for the median. For clarity in distributions with long tails, the violins are set to equal width. These distributions can be used to investigate the types of galaxy morphology found in each cluster.}
    \label{fig:distributions}
\end{figure*}

\autoref{fig:distributions} highlights the distribution of vote fractions for several key questions in the Galaxy Zoo decision tree, for galaxies in each cluster, as well as for the full labelled sample for comparison. To remove spurious components of the distributions, we set a rather stringent minimum threshold of 10 votes for each question. Using these distributions, we can investigate more deeply the types of galaxy residing in each cluster. 

Based on visual inspection of examples of cluster members, we excluded clusters 7, 8, 9, 11, 13, 15, 17 and 19 from the plots as they predominantly contained artefacts (some examples of these are shown in \autoref{fig:artefacts_cluster_samples}). Inevitably, some clusters (most notably cluster 19) contain some real and interesting sources among the artefacts, but these sources are in the minority. It should be noted that most of the artefacts in these clusters were introduced during preprocessing and were not present in the original data. While the preprocessing could certainly be improved to avoid the inclusion of these artefacts, artefacts are nonetheless often introduced in scientific pipelines and it is encouraging to see the clustering algorithm is able to trivially detect these. The number of galaxies excluded is 5332, out of a total of 65290 sources. We use the distributions of \autoref{fig:distributions} to better understand the types of galaxies found in each of the remaining clusters.

The first Galaxy Zoo decision tree question we considered is \emph{``Is the galaxy simply smooth and rounded, with no sign of a disk?''}. In the top panel of \autoref{fig:distributions}, we show the distributions for the fraction of votes for ``Features or disk'' (in other words, the galaxy is not smooth or an artefact). The distributions for clusters 1, 3, 12 and 14 strongly suggest they are largely smooth and probably elliptical galaxies. It is interesting to note that clusters 1 and 12 contain well-resolved sources while 3 and 14 are much lower resolution, which may result in the citizen scientists voting them to be smooth. Cluster 5 appears to consist almost entirely of spirals, while the remaining clusters have less clean distributions. Clusters 0 and 18 contains a mix of morphologies, but closer investigation of the question \emph{``Is there any sign of a spiral pattern?''} (not shown in \autoref{fig:distributions}), as well as inspection of the examples in \autoref{fig:clean_cluster_samples}, suggests that these groups consists largely of spirals.

Cluster 6 shows fairly broad distributions throughout, which reflects the fact that this cluster is not well localised in \autoref{fig:non_clean_cluster_samples}. Inspection of the examples of cluster members suggests that algorithm grouped together objects with nearby (usually coincident) sources, rather than primarily by morphology. Although not shown in order to reduce complexity of the figure, we also investigated the ``Artefact'' answer to the first question in the Galaxy Zoo decision tree. Only clusters 16 and 19 indicated a higher than average number of artefacts. We discarded cluster 19 based on visual inspection indicating that artefacts dominated this cluster. Cluster 16 however, consists of a mix of artefacts and galaxies with interesting morphology so we kept cluster 16 in the sample.

We further investigated these clusters by examining the answer to the question \emph{``How rounded is it?''}, focusing on the ``Cigar shaped'' galaxies in the second panel of \autoref{fig:distributions}. As this question in the decision tree only applies to galaxies that appear smooth, we only considered objects for which the fraction of votes for a smooth object is greater than 0.5. We also considered the question \emph{``Could this be a disk viewed edge-on?''} in the third panel. We applied a threshold of 0.5 to the fraction of votes that consider the galaxy to be featured. These two plots together show that clusters 2 and 10 both consist of elongated galaxies, but the algorithm is not able to easily distinguish between cigar-shaped ellipticals and edge-on spirals (although more cigar-shaped galaxies seem to occur in cluster 2 than 10). This is not surprising as edge-on spirals and elongated ellipticals can be indistinguishable even for a human if the inclination angle is high enough or the resolution low. Finally, we note that cluster 16 seems to have a mix of edge-on and face-on galaxies and is one of the more complex grouping of galaxies.

\begin{table*}
\centering
\begin{tabular}{ll}
\hline
Cluster number & Description \\
\hline
0 &  Predominantly spiral galaxies\\
1 &  Mostly ellipticals and some spirals with prominent bulges\\
2 &  Edge-on disks and some cigar-shaped ellipticals\\
3 &  Predominantly ellipticals but does contain some spirals\\
4 &  A large mix of morphologies, including some apparent mergers and some artefacts\\
5 &  Almost entirely spiral galaxies\\
6 &  Mixed morphology, dominated by galaxies that have a (sometimes coincident) companion\\
7 &  Artefacts\\
8 &  Artefacts\\
9 &  Artefacts\\
10 &  Edge-on disks and some cigar-shaped ellipticals\\
11 &  Artefacts\\
12 &  Predominantly ellipticals\\
13 &  Artefacts\\
14 &  Predominantly ellipticals\\
15 &  Artefacts\\
16 &  A large mix of morphologies, including some apparent mergers and some artefacts\\
17 &  Artefacts\\
18 &  Predominantly spiral galaxies\\
19 &  Artefacts\\
\hline
\end{tabular}
\caption{Qualitative summary of each of the clusters detected by applying a Bayesian Gaussian mixture model to the \gz{} data. \new{These qualitative conclusions are further supported by a quantitative analysis using the evaluation subset, detailed in the Appendix, \autoref{appendix:clusters}.}}
\label{tab:clusters}
\end{table*}

In the fourth panel of \autoref{fig:distributions}, we considered the question \emph{``How prominent is the central bulge, compared with the rest of the galaxy?''} and focused on the distributions for the answer ``Dominant bulge''. Similarly to the previous questions, we required the fraction of votes for the featured question to be greater than 0.5. This plot reveals that although clusters 12 and 14 consist of mostly ellipticals, they also include some disk galaxies with prominent bulges that appear similar to ellipticals. By viewing the examples in \autoref{fig:clean_cluster_samples} and \autoref{fig:non_clean_cluster_samples}, it is easy to see how these objects can be confused even by a human. Cluster 1, which contains some mixed morphologies, also seems to consist mostly of galaxies with a dominant bulge, although we note that this cluster contains a significant number of artefacts.

The final panel of \autoref{fig:distributions} focuses on the question \emph{``Is the galaxy merging or is there any sign of tidal debris?''} Because this question has multiple answers, we focused on the negative to look for sources that have any indication of mergers at all. We once again required at least 50\% of users to have voted that the galaxy is featured to include it in the sample. The plot suggests that clusters 1, 4 and 16 may have merger candidates, although they will be mixed with other sources. \autoref{fig:non_clean_cluster_samples} shows that clusters 1, 4 and 16 both appear to have complex, multiple sources, with cluster 16 having more ``zoomed out'' examples.

\autoref{tab:clusters} gives a qualitative description of the galaxies in each cluster, based on \autoref{fig:distributions} and Figures \ref{fig:clean_cluster_samples}-\ref{fig:artefacts_cluster_samples}.

\subsection{Cluster membership of evaluation subset}
In \autoref{sec:data_labels} we introduced a ``clean'' and simple evaluation set of labelled objects, consisting of round ellipticals, spirals and edge-on \km{galaxies}. We investigated which clusters these objects were assigned to:
\begin{itemize}
    \item Round ellipticals - 77.2\% found in clusters 12, 14, 3 and 1.
    \item Spirals - 79.3\% found in clusters 5, 18 and 0.
    \item Edge-on \km{galaxies} - 79.3\% found in clusters 10, 2 and 6.
\end{itemize}
\new{\autoref{appendix:clusters} in the Appendix shows this break down in more detail.} In all cases the majority of the sources were found in the first one or two clusters listed. These results broadly reflect the more qualitative conclusions drawn in \autoref{sec:clustering_results}. 

How can this be useful? By simply inspecting a few objects in each cluster, we can apply labels to the clusters as above. This simple approach results in an overall accuracy of 77\%\ for our evaluation subset. While this is clearly far below the performance of a supervised algorithm, it is impressively high for a fully unsupervised approach. This approach can be a highly efficient way to select and label samples for training downstream supervised learning algorithms, as well as being useful for removing artefacts and identifying unusual clusters of objects. 

\section{Anomaly detection}
\label{sec:anomaly_detection}
To further demonstrate the general utility of these self-supervised features, we applied anomaly detection to the dataset. Detecting rare and unusual objects in massive datasets is a key application of unsupervised learning. In order to compare with some ground truth labels and test performance, we focused on merger candidates as a specific type of anomalous object to search for. While it is completely possible to detect more rare types of sources with the same algorithm and features, we did not have easy access to labels of such sources, and considered the merger example as a suitable anomaly detection demonstration.

\begin{figure*}
    \centering
    \includegraphics[width = \linewidth]{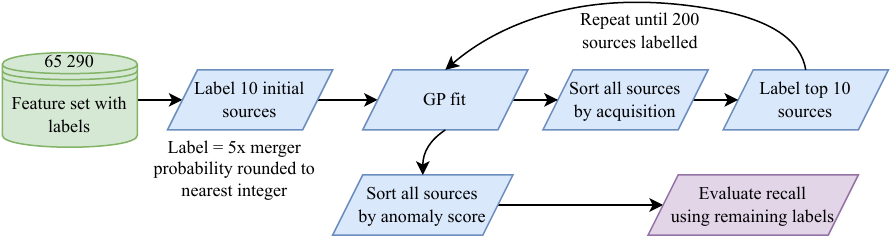}
    \caption{\new{Flow diagram of the anomaly detection section of our methodology. Features are represented as cylinders and processes as parallelograms. The number on top of the cylinder symbol indicate how many objects are in that dataset. }}
    \label{fig:flow_diagram_anomaly}
\end{figure*}

\subsection{Astronomaly}
We applied the \astronomaly{} framework \citep{Lochner2021}, with the alternative anomaly detection approach introduced in \citet{Walmsley_2022}, which we found works well for this dataset. \new{The procedure we follow is illustrated in the flow diagram of \autoref{fig:flow_diagram_anomaly}.} \astronomaly{} employs active learning to rapidly identify not just anomalies in data, but anomalies that are specifically of interest to the user. The key concept of \astronomaly{} is that ``blind'' anomaly detection, using machine learning, will usually detect a large number of sources which, while anomalous, are not of interest to a scientist. Thus, \astronomaly{} employs an interactive framework to obtain very few, strategic labels from a user in order to improve the detection algorithm. This is somewhat similar to recommendation engines used in popular music or video streaming services. In our case, uninteresting anomalies would include artefacts or galaxies with unusual morphologies, while anomalies of interest would be merger candidates. \astronomaly{} requires the user to score, on a scale of 0 to 5, examples of objects according to their ``interestingness''. In machine learning nomenclature, the algorithm requests these labels from an ``oracle''. Normally, this oracle would be a human expert who labels objects according to how interesting they are for the science interests of that expert. However, we had access to a set of ground truth labels from the citizen scientist volunteers to provide labels when requested by the algorithm. Although we have access to a large number of ground truth labels, only a subset of these were provided to the active learning algorithm. 

We made use of the Galaxy Zoo decision tree question \emph{``Is the galaxy merging or is there any sign of tidal debris?''}. This question has four possible answers: ``Merging'', ``Tidal debris'', ``Both'' or ``Neither''. Visual inspection suggested that galaxies with a high voter fraction for ``Tidal debris'' are only mildly disturbed so we added the voter fractions for ``Merging'' and ``Both'' and took that number to be the probability that an image is a merging system. We scaled this probability to lie on the range of 0-5 and rounded to the nearest integer to simulate \astronomaly{}'s scoring mechanism.  
\begin{figure}
    \centering
    \includegraphics[width = 0.95\linewidth]{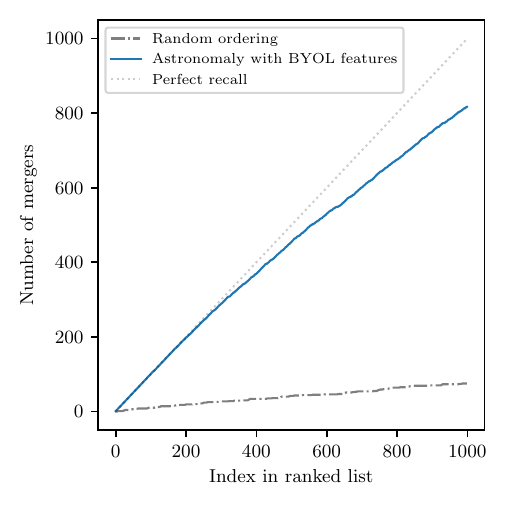}
    \caption{Number of mergers detected as a function of index in a list of sources ordered by anomaly score, after active learning is applied to 200 sources. The theoretical performance of a random order and perfect recall is shown for reference. \astronomaly{} is able to efficiently recover merger candidates without requiring significant labelling.}
    \label{fig:anomaly_recall}
\end{figure}
The key question is, what threshold do we apply to decide if an image corresponds to a true positive merger? It seems overall voter confidence in this question was low since a high threshold (like 0.8) cuts out a large number of obvious merger candidates. We follow the methodology of \citet{gz_for_merger} and consider sources with a threshold of 0.4 to be very likely merger candidates. This will naturally not result in a very clean sample, but visual inspection suggests this cut is appropriate. We also ensure the voter fraction is at least 10 before considering a source a ``true'' merger. This results in 4990 mergers or 7\% of the dataset considered as interesting anomalies.

We passed the features described in \autoref{sec:methodology} after applying PCA to \astronomaly{}. The active learning approach of \citet{Walmsley_2022} requires an initial sample of objects with labels. While \citet{Walmsley_2022} used a random sample of galaxies to begin the process, we found that this introduces a large degree of randomness in the results, especially with small numbers of labels. Instead, in order to find a deterministic sample of galaxies that should broadly span the feature space, we sorted the features by their first principal component and selected an example from every tenth percentile (i.e. 10 objects equally spaced by index in the sorted list, not feature value). We labelled these with an integer score from 0 to 5, based on the merger probability.

This initial set of labels allowed us to proceed to the active learning step of the \astronomaly{} pipeline. The goal is to train a regression algorithm to learn, with as few examples as possible, to predict the ``interestingness'' score of every object in the dataset thus quickly finding the interesting anomalies. As in \citet{Walmsley_2022}, we use a Gaussian Process (GP) to perform the regression. \new{A GP estimates a probability distribution of functions in order to predict a mean function value, as well as its corresponding uncertainty, for any input. We use this to predict the ``interestingness'' score at any given set of features, based on the small number of labels from the user.} The space of possible functions is given by the kernel. We use a Mat\'ern kernel, added to a white noise kernel to model intrinsic noise in the labelling, and allow the package \textsc{scikit-learn} to automatically optimise kernel hyperparameters.

\begin{figure*}
    \centering
    \includegraphics[width = 0.95\linewidth]{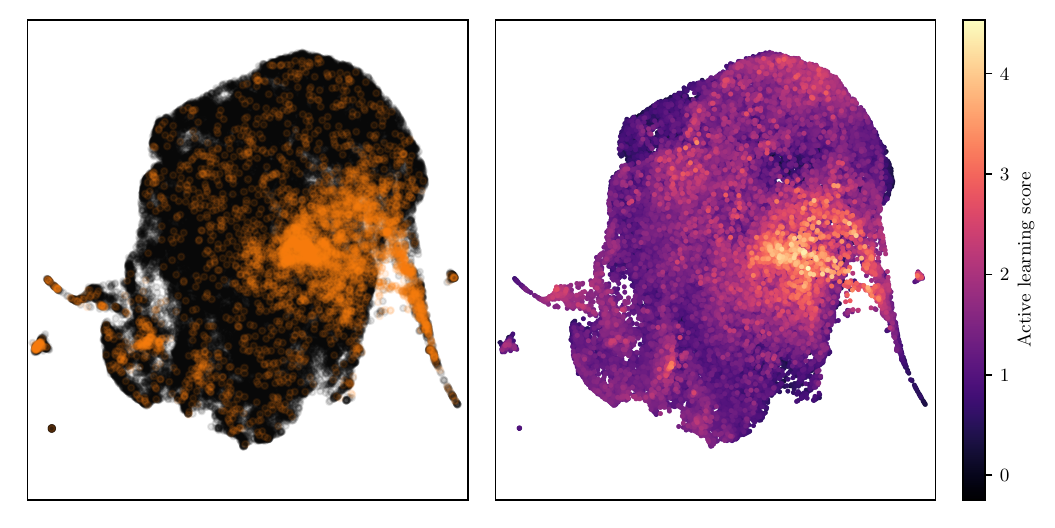}
    \caption{UMAP plots of the same \gz{} feature space highlighting the ``ground truth'' mergers (left) and showing the anomaly score after applying active learning (right). It is clear that \astronomaly{}, after training on just 200 examples, effectively prioritises the regions of feature space where mergers are likely to be found.}
    \label{fig:umap_mergers}
\end{figure*}

One of the most important advantages of a GP is that it calculates the uncertainty in its estimates, giving an indication of which regions of feature space are poorly constrained. This can be used in an acquisition function, which allows optimal selection of targets for the oracle to label in the next iteration that will improve the algorithm. We used exactly the same acquisition function as \citet{Walmsley_2022}, which is the maximum expected improvement. This function balances improving the score estimate across feature space by prioritising regions of high uncertainty and honing in on specific regions known to contain many high scoring objects. It has one tuning parameter, the trade-off $\eta$ which we set to 0.5. This tends to prioritise known interesting regions over exploration, resulting in quickly finding mergers. We found that changing the $\eta$ parameter did not strongly impact our results.

To perform our merger search, we labelled an initial 10 sources as described above and then trained the GP and labelled the top 10 unlabelled sources with the highest acquisition function. We repeated this process until 200 sources were labelled and then sorted the entire dataset according to estimated ``interestingness'' score. This corresponds to labelling just 0.3\% of the data.

In general, \textsc{Astronomaly} does not classify sources as anomalous or not, rather sorting the list and allowing a user to decide how far down the list they are willing to explore in order to look for interesting sources. We can then consider the recall, in terms of number of interesting anomalies detected, as a function of index in the ranked list to determine the effectiveness of the anomaly detection algorithm, shown in 
\autoref{fig:anomaly_recall}. The anomaly detection algorithm very efficiently recovers a large sample of mergers within the first few hundred sources with a high anomaly score. This excellent performance should obviously be taken with some caution as the threshold to be considered a merger is fairly low.

The two UMAP plots in \autoref{fig:umap_mergers} demonstrate how the active learning score successfully maps the overdensity of mergers in feature space. While some mergers are buried among other, non-anomalous, sources, the majority are grouped together again highlighting that the self-supervised features effectively groups sources with visually similar morphology.

\begin{figure*}
    \centering
    \includegraphics[width = 0.95\linewidth]{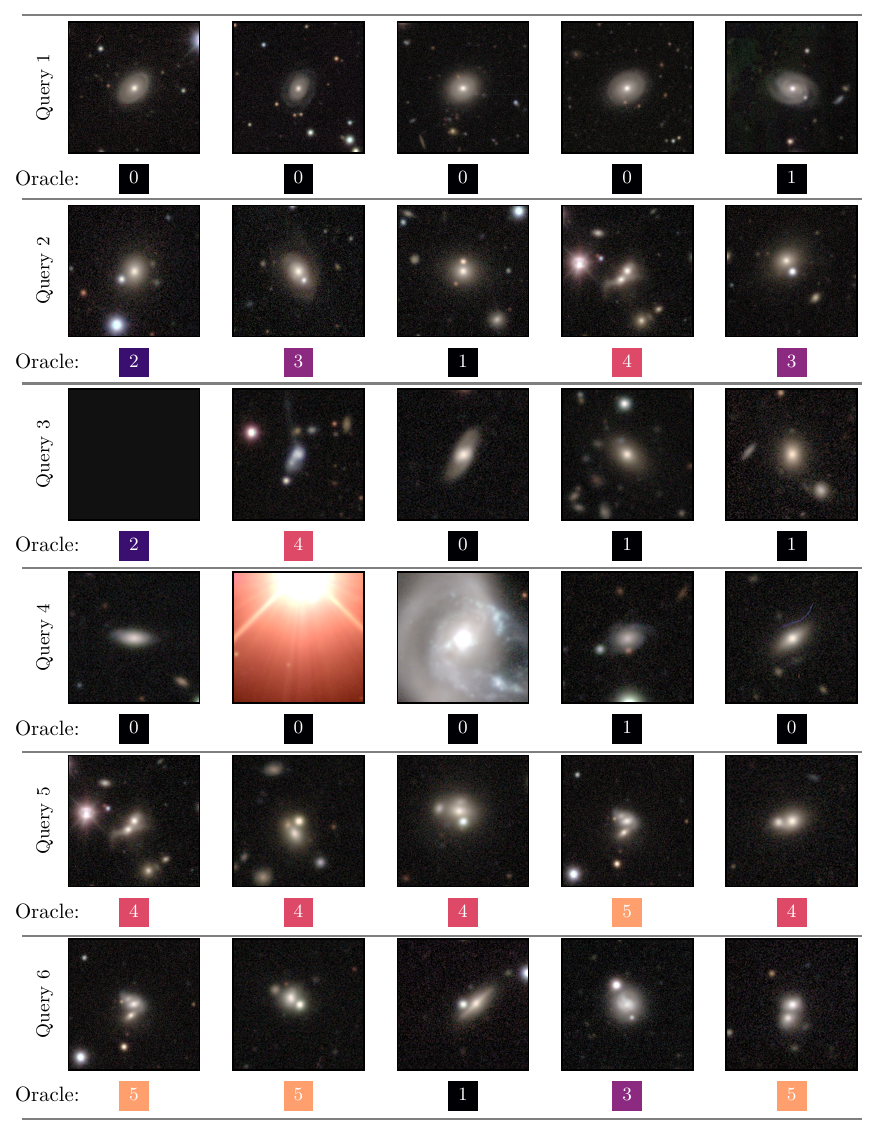}
    \caption{An illustration of the active learning approach used to detect anomalies (in this case merger candidates). Each iteration (query) consists of 10 examples being shown to the ``oracle'' after which the algorithm is retrained, ordered by acquisition score, and the next 10 objects are shown to the oracle. Only the top 5 examples for each query are shown here due to space considerations. The score the oracle gives (in this case, the labels are given from the citizen scientist votes) are shown in a box under each image. It can be seen that the algorithm quickly moves away from the more boring sources and hones in on merger candidates after seeing a handful of examples. Note that the blank image in query 3 is actually a completely black image in the \gz{} dataset.}
    \label{fig:example_mergers}
\end{figure*}

\autoref{fig:example_mergers} demonstrates that the majority of objects given a high score are indeed merger candidates. This figure attempts to demonstrate the active learning process by displaying the first five sources shown to the oracle each iteration, ordered by acquisition score (i.e. the five sources assigned the highest priority for labelling). After a few rounds querying some uninteresting objects, the algorithm quickly hones in on merger candidates which dominate the sources as quickly as the fifth query (40 sources labelled). 

\subsection{A similarity search for ring galaxies}
Inspired by \citep{Walmsley_2022} and other recent works, we perform a similarity search for a particularly unusual type of source, in this case choosing a few candidate ring galaxies. These objects are truly rare and we could use \astronomaly{} to perform a dedicated search for them, however we have no ground truth labels to compare against so we instead opted to demonstrate the utility of the self-supervised features using a similarity search. We select two example ring galaxies and use a simple Euclidean search on the PCA-reduced feature space to find the eight nearest neighbours to each source. 

\begin{figure}
    \centering
    \includegraphics[width = \linewidth]{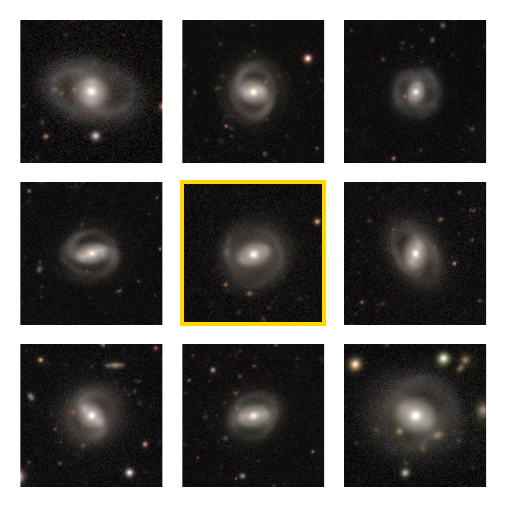}
    \caption{Similarity search for ring galaxies. The eight sources surrounding the central source are its eight nearest neighbours in feature space, according to the Euclidean distance.}
    \label{fig:similarity1}
\end{figure}

\begin{figure}
    \centering
    \includegraphics[width = \linewidth]{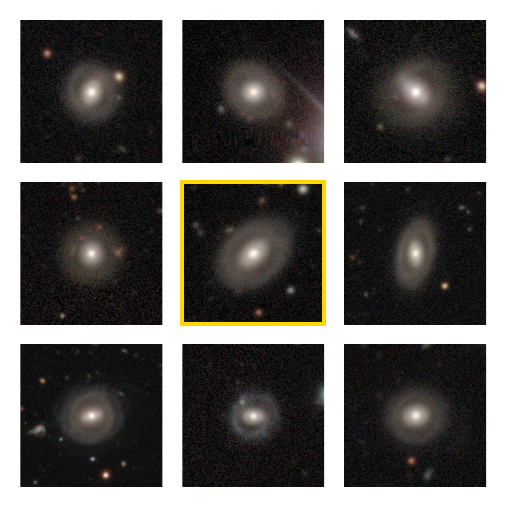}
    \caption{Same as \autoref{fig:similarity1}, we highlight a specific ring galaxy (central source) and its eight nearest neighbours in feature space.}
    \label{fig:similarity2}
\end{figure}

\autoref{fig:similarity1} and \autoref{fig:similarity2} show the two selected sources and their eight nearest neighbours. It is clear that they share very similar morphology, indicating that these features can be useful for rapidly identifying samples of rare objects as soon as an initial example is discovered.

\section{Application to a radio dataset}
\label{sec:mirabest}
As a final application to test the general utility of self-supervised learning for feature extraction in astronomy, we turn to a completely different type of dataset and apply the exact same approach.

\subsection{MiraBest data}
\begin{figure}
    \centering
    \includegraphics[width = \linewidth]{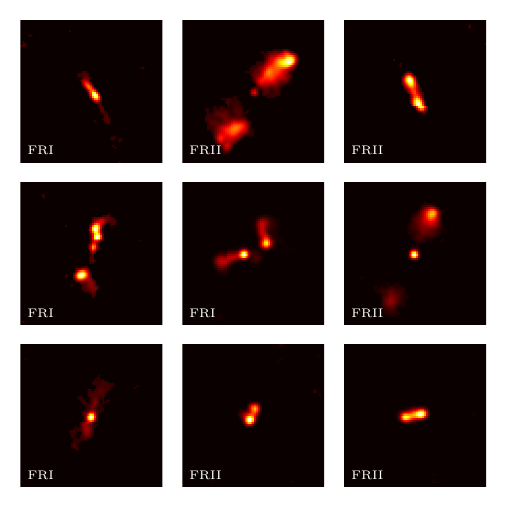}
    \caption{A random set of radio galaxies from the MiraBest training sample with the true label reported in the corner.}
    \label{fig:mirabest_examples}
\end{figure}

We chose the radio image dataset MiraBest \citep{Porter2023}, a set of 1256 labelled radio images\footnote{Available at \url{https://zenodo.org/records/5588282}.} from the Karl G. Jansky Very Large Array\footnote{\url{https://science.nrao.edu/facilities/vla/}}. Originally classified in \citet{Miraghaei2017}, this sample of radio-loud active galactic nuclei consists of Fanaroff-Riley I and II sources, as well as some sources that appear to be hybrid between the two classes. The Fanaroff-Riley (FR) dichotomy, first proposed in \citet{Fanaroff1974}, is a morphological classification based on the distance of the brightest regions of the radio jets from the core. FRI galaxies become fainter towards the edge and tend to have a lower luminosity in general, while FRII galaxies are edge-brightened and more luminous. This is the most commonly used morphological classification in radio astronomy, although there is a high degree of variation in source morphology and many examples of sources that are challenging to classify. A random sample of FRI and FRII galaxies from the MiraBest dataset is shown in \autoref{fig:mirabest_examples}.

We chose to apply our techniques to the MiraBest dataset for three reasons:
\begin{enumerate}
    \item Radio images differ substantially from optical images, providing a good stress test for BYOL.
    \item There are very few labelled radio image datasets and MiraBest has the added advantage of distinguishing between confidently and unconfidently labelled sources. We could thus train BYOL on the full dataset but use only the confidently-labelled sources to evaluate.
    \item With new radio telescopes such as MeerKAT \citep{MeerKAT}, ASKAP \citep{ASKAP} and LOFAR \citep{LOFAR} pushing into new regimes of sensitivity, resolution and sky coverage, radio astronomy is a field where unsupervised learning is expected to prove extremely useful.
\end{enumerate}
As the MiraBest sources were already cropped to some degree, we did not perform any resizing operation. We ran BYOL on this dataset using the exact same hyperparameters described in \autoref{sec:methodology} except that we set the batch size to 32 (owing to the smaller dataset) and the number of epochs to 50, since we found that it took longer to produce useful representations. Even with the increased number of epochs, the small dataset size means we are able to train BYOL in just 11 minutes, using the same GPU mentioned in \autoref{sec:augmentations}.

This dataset represents an interesting opportunity to test which is the best set of weights to initialise the network with. The idea is that so-called ``foundation models'', already trained on large datasets, can then be easily fine-tuned using BYOL on smaller, different datasets for downstream tasks. We explore four scenarios for the weights initialisation:
\begin{enumerate}
\renewcommand{\labelenumi}{(\arabic{enumi})}
    \item Random weights
    \item ImageNet trained weights
    \item \gz{} trained weights
    \item Radio Galaxy Zoo trained weights
\end{enumerate}
The last item on the list is derived from \citet{slijepcevic2023radio} where BYOL was trained on a sample of non-public radio images. Fortunately, the authors made their trained model available and the architecture is similar to the model we used. We were able to use this model and fine-tune it to MiraBest. It is worth noting that this was also done in \citet{slijepcevic2023radio} but there are some differences in approach, resulting in somewhat different accuracies achieved for the confident MiraBest sample.

\subsection{Evaluation}
\begin{figure}
    \centering
    \includegraphics[width = \linewidth]{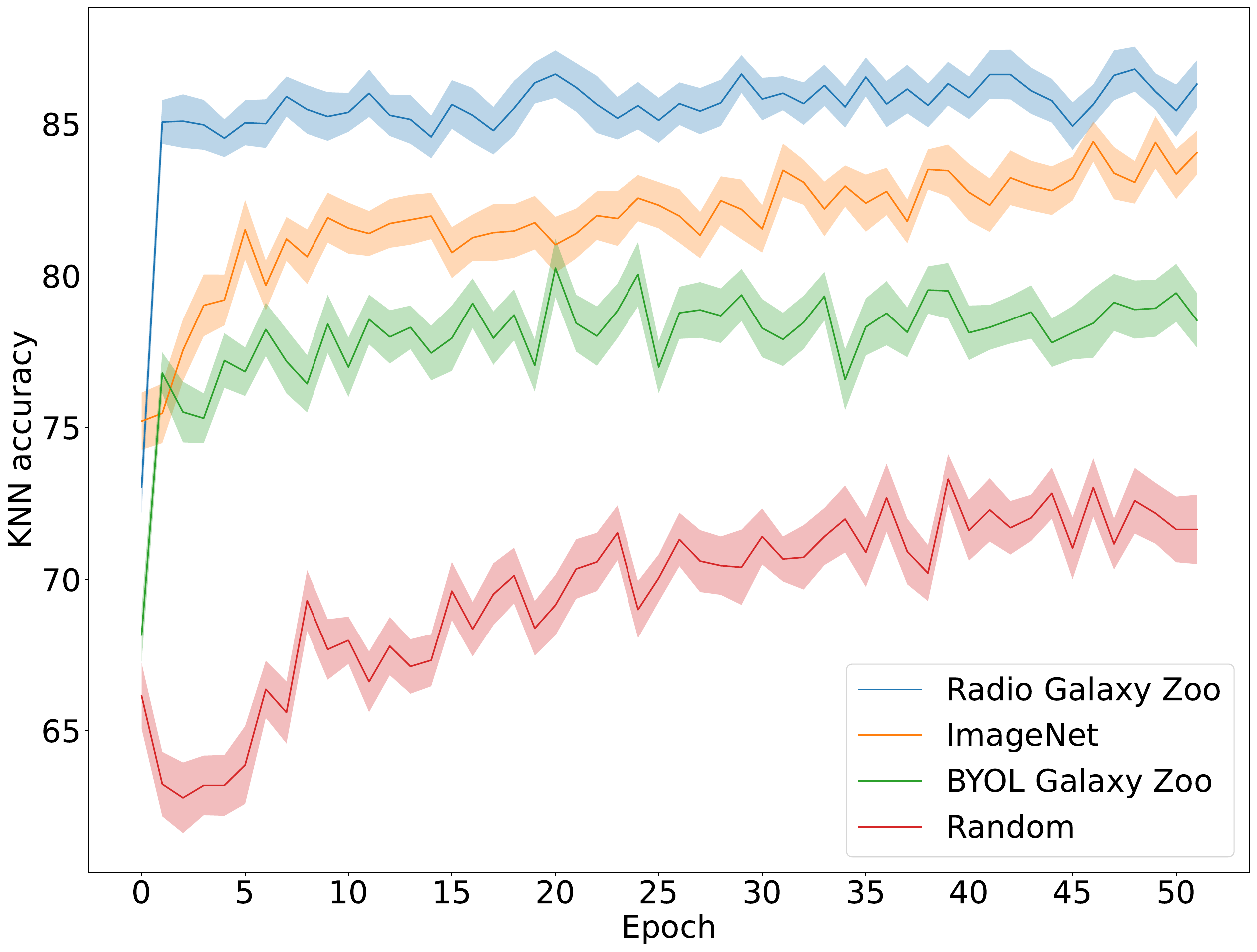}
    \caption{KNN accuracy as a function of epoch for the  confidently-labelled MiraBest dataset for each of the four scenarios with different initialisation of the weights. It is clear that fine-tuning is superior to random-initialisation. Using a network trained on very similar data provides the highest performance but this can be almost matched by using a network trained a large dataset of terrestrial images (ImageNet). \new{The corresponding F1 scores (harmonic mean of the precision and recall) are listed in \autoref{tab:f1_mirabest} in the Appendix.}}
    \label{fig:knn_accuracy_mirabest}
\end{figure}

For each of the four scenarios, we trained a KNN on the confidently labelled data, used a 75\%-25\%\ training-test split and ran the classifier for 50 different random splits at each epoch. \autoref{fig:knn_accuracy_mirabest} shows the accuracy as a function of epoch for each scenario. Firstly, it is clear that any of the three types of fine-tuning dramatically outperforms random weights, which tends instead towards a very poor representation in the limited number of epochs and with such a small training set. It is not surprising that fine-tuning from the Radio Galaxy Zoo network, trained using very similar data, outperforms all other options. The overall accuracy is comparable to that in \citet{slijepcevic2023radio}, if slightly lower because we are using KNN rather than a full CNN for the final classification step. 

What is perhaps more interesting is that fine-tuning an ImageNet-trained network outperforms that of our Galaxy Zoo-trained network. This is counter-intuitive since radio galaxies are more similar to optical galaxies than terrestrial images of animals and objects. The reason for this performance difference could be because the ImageNet classification task involves of thousands of classes, forcing the network to learn very general representations (as was found in \citet{Walmsley_2022}) or it may be simply due to the size of the training sets, with ImageNet being an order of magnitude larger than the \gz{} sample we used. Either explanation bodes well for the general use of foundation models in astronomy as it suggests that a model trained on a completely different, but large, dataset can still be useful as a feature extractor if fine-tuned to the dataset it's being applied to.

This conclusion is further supported by examining the feature space of each of the four initialisations described above. \autoref{fig:mirabest_feature_space} shows the UMAP plot for each case for only the confident MiraBest sample, with FRI and FRII galaxies indicated with different markers. No separation between classes is seen when random weights are used, while the Radio Galaxy Zoo and ImageNet weights show fairly significant division between classes. 

\subsection{Clustering}
\label{sec:mirabest_clustering}
\begin{figure*}
    \centering
    \includegraphics[width = \linewidth]{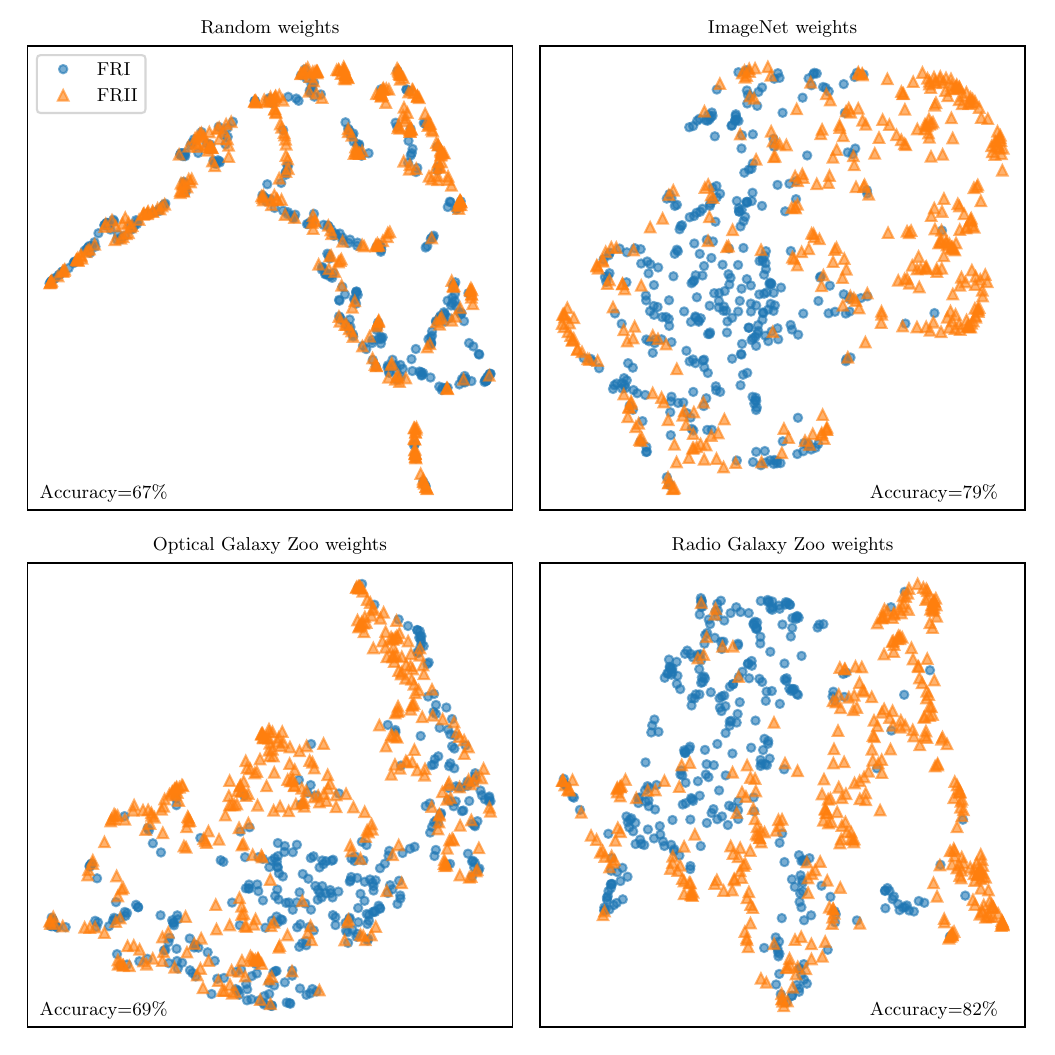}
    \caption{UMAP plots for four different sets of initial weights: random (top left), ImageNet (top right), our Galaxy Zoo-trained BYOL (bottom left) and the BYOL trained on Radio Galaxy Zoo from \citet{slijepcevic2023radio} (bottom right). The features are shown for the confident MiraBest sample only and FRI and FRII galaxies are indicated with different symbols. When clustering is applied, if we assign an entire cluster with the same class as the majority source in that cluster, we obtain the accuracy indicated in each figure. This accuracy is close to that of the supervised approach shown in \autoref{fig:knn_accuracy_mirabest}, indicating that BYOL successfully learns features useful for both clustering and supervised classification. Finally, both the accuracy and the visual appearance of the UMAP plots seem to suggest that a large training set is an important factor when fine-tuning a pretrained network, as the ImageNet weights outperform the optical Galaxy Zoo weights despite the fact that they are derived from a dataset less similar to MiraBest.}
    \label{fig:mirabest_feature_space}
\end{figure*}

Despite the small sample size, we managed to apply the same clustering approach as described in \autoref{sec:clustering}. We first applied PCA to each of the four sets of features with a threshold explained variance of 95\%, resulting in reduced features of size 3, 31, 58 and 16 for the random, ImageNet, Galaxy Zoo and Radio Galaxy Zoo weights respectively. We then applied a Bayesian Gaussian mixture model with up to 20 components to each set of reduced features. To simulate a real unsupervised learning application, we examine each cluster to check which is the dominant source in each. The cluster is then labelled as that source (i.e. if more than 50\% of the objects in the cluster are FRI, all objects in the cluster are labelled FRI). While in this case we know the ground truth which makes the labelling easier, in a truly unlabelled dataset one would have to simply investigate a few samples and decide which is the dominant source for that cluster. By applying these labels, we can compute the overall accuracy of this pseudo-classifier, which is reported in \autoref{fig:mirabest_feature_space}. Remarkably, we obtain almost as high accuracy as when we use a traditional KNN approach. This implies that a fully unsupervised approach, fine-tuning a foundation model and applying clustering, can at a minimum quickly produce potential training sets that can later by optimised through minimal human inspection and labelling.

\section{Conclusions}
\label{sec:conclusions}
Inspired by a rapid increase in size and complexity of astronomical datasets, we have explored an array of unsupervised learning methods for data exploration and scientific discovery. Deep learning has proven itself in the arena of supervised learning, but recently has begun to be used to obtain useful representations of high-dimensional data, such as images, without requiring expensive human-provided labels. We have shown that self-supervised learning, specifically the non-contrastive algorithm Bootstrap Your Own Latent (BYOL), is highly effective at feature extraction for astronomical images. We applied this algorithm to two quite different datasets: a sample of hundreds of thousands of optical images from \gz{} and a relatively small set of radio galaxies from MiraBest. While the latent features perform well for a set of supervised learning tasks, the primary goal was to generate features suitable for unsupervised learning tasks, including clustering and anomaly detection.

As with any deep learning application, we found that preprocessing choices are important. Our procedure to crop and resize each image, removing biases with respect to object size, inadvertently introduced artefacts in the data. However, we showed that these can be largely removed by applying an initial pretrained CNN as a feature extractor, visualising the space using the manifold embedding technique UMAP and cutting out an obvious large cluster of anomalies. This simple methodology could prove useful for very quickly identifying artefacts in other datasets as it requires no training at all. However, since the network was trained on a completely different dataset (ImageNet), it was not an appropriate feature extractor for more subtle distinctions between morphology.

By instead using the self-supervised learning algorithm BYOL, we were able to extract features for both the optical and radio dataset that performed well when combined with a k-nearest neighbours (KNN) algorithm to classify a subset of sources with known classes. While this approach is unlikely to compete with a purpose-built CNN when a large training set is available, it does demonstrate that BYOL is learning a useful representation of the data. 

We used the same features for unsupervised learning applications, finding that a Bayesian Gaussian mixture model (BGMM) was able to successfully detect clusters of galaxies with similar morphologies for both datasets. For the \gz{} dataset, we analysed the citizen scientist votes to determine the types of sources found in each cluster. While at times the algorithm places sources of different morphology in the same cluster (usually when confused by the presence of a coincident source or artefact), many clusters are relatively pure. For the MiraBest case, the accuracy of the fully unsupervised clustering method is close to that of a supervised KNN classifier. 

This approach could be useful for quickly building initial training sets for subsequent supervised learning algorithms, which is currently a laborious process. While we did not explore this application here, one could imagine using the probability of a particular source belonging to a cluster (which is given by BGMM) to determine whether to include the source in the training set, or request an oracle (i.e. an expert or citizen scientist) to identify the source. A clustering analysis could also reveal unexpected groupings of sources in large, deep surveys.

For the \gz{} dataset, we also applied the \astronomaly{} framework to detect anomalies. The only Galaxy Zoo decision tree category we had access to which could be considered anomalous was that of mergers, so we aimed to use anomaly detection to find merger candidates. We found that \astronomaly{} was able to use the BYOL-trained features to efficiently locate merger candidates, although this was using a relatively low cut on probability of being considered a merger (which was the same cut used in \citet{gz_for_merger}). We also showed how a similarity search can quickly identify candidate ring galaxies, another rare type of source. While unsupervised methods generally will not outperform a trained classifier, for cases where training data is spare or the object in question is rare, the BYOL-derived features are clearly effective at locating objects of interest.

We found that fine-tuning the BYOL algorithm from the weights of a pretrained network far outperforms the more common approach of initialising the weights randomly. Interestingly, for the MiraBest data we found that ImageNet weights outperformed those from BYOL trained on \gz{}, despite the optical dataset being far more similar to the radio than the terrestrial one. This could be due to the much larger dataset when performing the initial training or the fact that the network in question was trained to classify a large number of diverse classes. This question could be answered by initialising the weights from a network trained with BYOL or another self-supervised learning method, rather than supervised learning, to identify which approach produces the most useful starting weights. Less surprisingly, we found a network trained (using BYOL) on a large set of radio galaxies outperformed all others when fine-tuned on MiraBest. This bodes well for the idea of providing foundation models, trained on large datasets, for astronomical data which can then be fine-tuned to any specific dataset and used for downstream tasks.

The key challenge in applying unsupervised methods to image data is usually feature extraction. This work demonstrates that this challenge can be overcome by leveraging self-supervised methods to extract meaningful representations of the data. These representations enable automated clustering, which in turn allows for a rapid removal of artefacts, a faster route to labelled training sets and the potential for discovering new patterns in the data. The same representations can also be used to find rare sources and even discover new classes of objects with minimal human intervention.

\section*{Acknowledgements}

ML and KM acknowledge support from the South African Radio Astronomy Observatory and the National Research Foundation (NRF) towards this research. Opinions expressed and conclusions arrived at, are those of the authors and are not necessarily to be attributed to the NRF. 

Author contribution statements: KM lead the research and development, including conceptualisation of many of the key ideas, and produced an early draft of the paper. ML supervised the project, performed the analysis in \autoref{sec:clustering_results},  \autoref{sec:anomaly_detection} and \autoref{sec:mirabest_clustering} and wrote the majority of the paper.

This research made use of the python programming language and the following open source packages: Numpy, SciPy \citep{Jones2001}, Matplotlib \citep{Hunter2007}, Seaborn \citep{Waskom2021}, scikit-learn \citep{scikit-learn}, Pandas \citep{McKinney2010}, Astropy \citep{Astropy1, Price-Whelan_2018}, umap-learn \citep{sainburg2021}, PyTorch \citep{pytorch} and BYOL-PyTorch \citep{chen2020}. The authors also acknowledge the surprising helpfulness of ChatGPT\footnote{\url{https://chat.openai.com/}} in coming up with ideas for the title based on the (human-written) abstract. It was not used for any other aspect of the paper.

We acknowledge the use of the ilifu cloud computing facility – www.ilifu.ac.za, a partnership between the University of Cape Town, the University of the Western Cape, the University of Stellenbosch, Sol Plaatje University and the Cape Peninsula University of Technology. The Ilifu facility is supported by contributions from the Inter-University Institute for Data Intensive Astronomy (IDIA – a partnership between the University of Cape Town, the University of Pretoria and the University of the Western Cape, the Computational Biology division at UCT and the Data Intensive Research Initiative of South Africa (DIRISA).

This publication uses data generated via the Zooniverse.org platform, development of which is funded by generous support, including a Global Impact Award from Google, and by a grant from the Alfred P. Sloan Foundation.

The Legacy Surveys consist of three individual and complementary projects: the Dark Energy Camera Legacy Survey (DECaLS; Proposal ID \#2014B-0404; PIs: David Schlegel and Arjun Dey), the Beijing-Arizona Sky Survey (BASS; NOAO Prop. ID \#2015A-0801; PIs: Zhou Xu and Xiaohui Fan), and the Mayall $z$-band Legacy Survey (MzLS; Prop. ID \#2016A-0453; PI: Arjun Dey). DECaLS, BASS and MzLS together include data obtained, respectively, at the Blanco telescope, Cerro Tololo Inter-American Observatory, NSF’s NOIRLab; the Bok telescope, Steward Observatory, University of Arizona; and the Mayall telescope, Kitt Peak National Observatory, NOIRLab. Pipeline processing and analyses of the data were supported by NOIRLab and the Lawrence Berkeley National Laboratory (LBNL). The Legacy Surveys project is honoured to be permitted to conduct astronomical research on Iolkam Du’ag (Kitt Peak), a mountain with particular significance to the Tohono O’odham Nation. NOIRLab is operated by the Association of Universities for Research in Astronomy (AURA) under a cooperative agreement with the National Science Foundation. LBNL is managed by the Regents of the University of California under contract to the U.S. Department of Energy. This project used data obtained with the Dark Energy Camera (DECam), which was constructed by the Dark Energy Survey (DES) collaboration. Funding for the DES Projects has been provided by the U.S. Department of Energy, the U.S. National Science Foundation, the Ministry of Science and Education of Spain, the Science and Technology Facilities Council of the United Kingdom, the Higher Education Funding Council for England, the National Center for Supercomputing Applications at the University of Illinois at Urbana-Champaign, the Kavli Institute of Cosmological Physics at the University of Chicago, Center for Cosmology and Astro-Particle Physics at the Ohio State University, the Mitchell Institute for Fundamental Physics and Astronomy at Texas A\&M University, Financiadora de Estudos e Projetos, Fundacao Carlos Chagas Filho de Amparo, Financiadora de Estudos e Projetos, Fundacao Carlos Chagas Filho de Amparo a Pesquisa do Estado do Rio de Janeiro, Conselho Nacional de Desenvolvimento Cientifico e Tecnologico and the Ministerio da Ciencia, Tecnologia e Inovacao, the Deutsche Forschungsgemeinschaft and the Collaborating Institutions in the Dark Energy Survey. The Collaborating Institutions are Argonne National Laboratory, the University of California at Santa Cruz, the University of Cambridge, Centro de Investigaciones Energeticas, Medioambientales y Tecnologicas-Madrid, the University of Chicago, University College London, the DES-Brazil Consortium, the University of Edinburgh, the Eidgenossische Technische Hochschule (ETH) Zurich, Fermi National Accelerator Laboratory, the University of Illinois at Urbana-Champaign, the Institut de Ciencies de l’Espai (IEEC/CSIC), the Institut de Fisica d’Altes Energies, Lawrence Berkeley National Laboratory, the Ludwig Maximilians Universitat Munchen and the associated Excellence Cluster Universe, the University of Michigan, NSF’s NOIRLab, the University of Nottingham, the Ohio State University, the University of Pennsylvania, the University of Portsmouth, SLAC National Accelerator Laboratory, Stanford University, the University of Sussex, and Texas A\&M University. BASS is a key project of the Telescope Access Program (TAP), which has been funded by the National Astronomical Observatories of China, the Chinese Academy of Sciences (the Strategic Priority Research Program “The Emergence of Cosmological Structures” Grant \# XDB09000000), and the Special Fund for Astronomy from the Ministry of Finance. The BASS is also supported by the External Cooperation Program of Chinese Academy of Sciences (Grant \# 114A11KYSB20160057), and Chinese National Natural Science Foundation (Grant \# 12120101003, \# 11433005). The Legacy Survey team makes use of data products from the Near-Earth Object Wide-field Infrared Survey Explorer (NEOWISE), which is a project of the Jet Propulsion Laboratory/California Institute of Technology. NEOWISE is funded by the National Aeronautics and Space Administration. The Legacy Surveys imaging of the DESI footprint is supported by the Director, Office of Science, Office of High Energy Physics of the U.S. Department of Energy under Contract No. DE-AC02-05CH1123, by the National Energy Research Scientific Computing Center, a DOE Office of Science User Facility under the same contract; and by the U.S. National Science Foundation, Division of Astronomical Sciences under Contract No. AST-0950945 to NOAO.

\section*{Data Availability}
The data used in this paper are publicly available at \url{https://zenodo.org/record/4573248} for the \gz{} data and \url{https://zenodo.org/records/5588282} for the MiraBest data.




\bibliographystyle{mnras}
\bibliography{refs} 

\appendix

\section{UMAP hyperparameters}
\label{appendix:umap}
\begin{figure*}
    \centering
    \includegraphics[width = 0.85\linewidth]{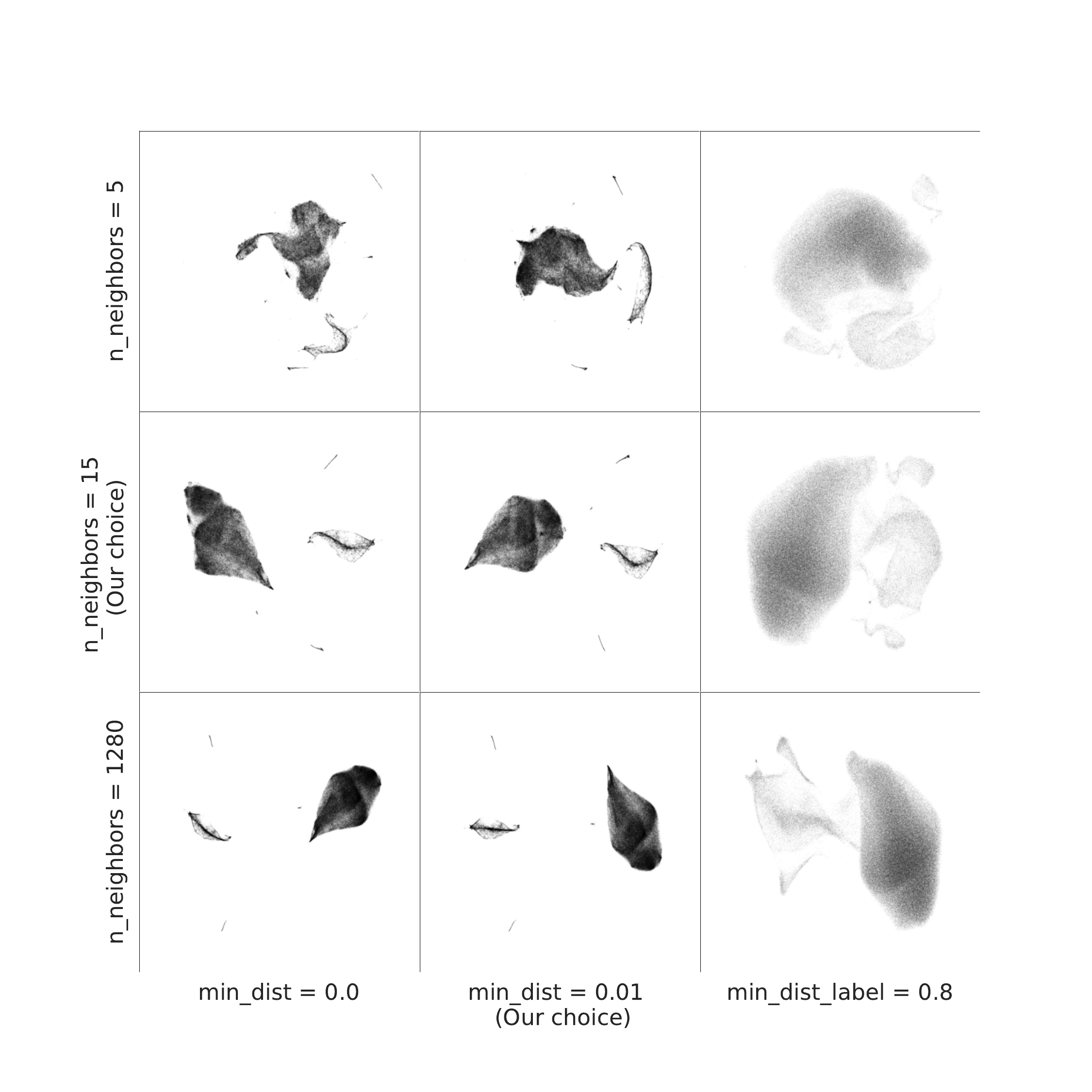}
    \caption{\km{UMAP produces different embeddings depending on the choices of the hyperparameters ``number of neighbours'' and ``minimum distance''. Our chosen UMAP embedding shows structures that are present for larger parameter values of ``number of neighbours'' and ``minimum distance'', and are therefore not artefacts of UMAP.}}
    \label{fig:umap_parameters}
\end{figure*}

\new{UMAP has several hyperparameters, the effects of which are discussed in \citet{umap}. In \autoref{fig:umap_parameters}, we show the effect, for the Galaxy Zoo dataset, of varying these parameters over a very broad range. Based on visual inspection, we set ``number of neighbours'' to 15, and the parameter ``minimum distance'' to 0.01 for our analysis. The plots are not sensitive to moderate variations in these parameters but large deviations such as shown here can negatively impact the structure and interpretation of UMAP plots.}

\section{Artefacts}
\label{appendix:artefacts}
\begin{figure}
 
     \begin{subfigure}{\linewidth}
         \centering
    \includegraphics[width = \linewidth]{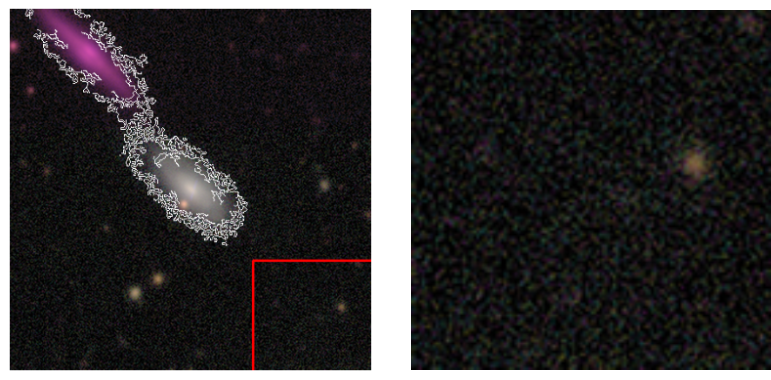}
         \caption{Artefact detected in Cluster 13}
         \label{fig: Artefact_cluster_13}
     \end{subfigure}
     \hfill
     \begin{subfigure}{\linewidth}
         \centering
    \includegraphics[width = \linewidth]{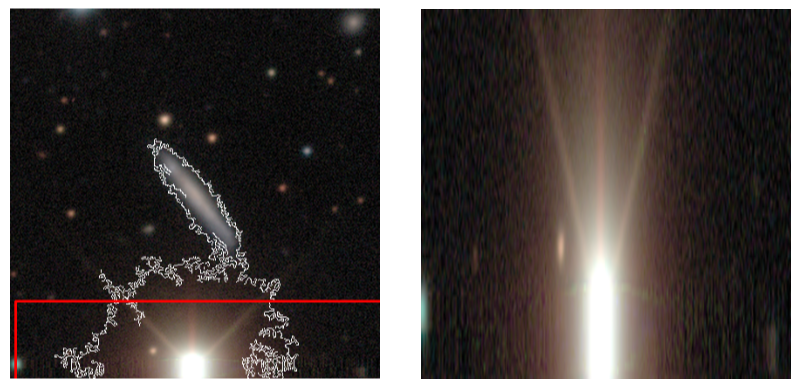}
         \caption{Artefact detected in Cluster 19}
         \label{fig: Artefact_cluster_19}
     \end{subfigure}
         \hfill
     \begin{subfigure}{\linewidth}
         \centering
    \includegraphics[width = \linewidth]{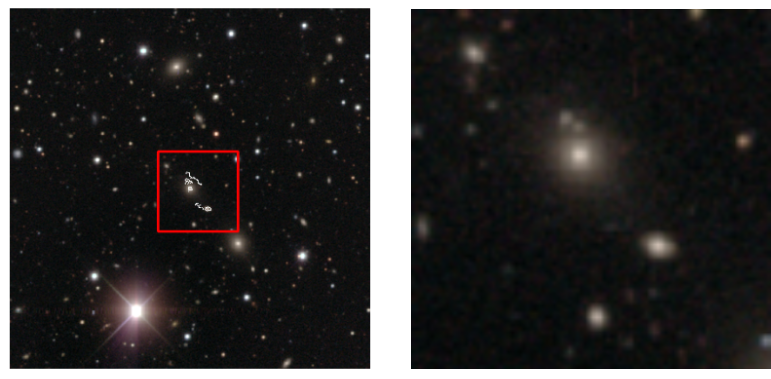}
         \caption{Artefact detected in Cluster 19}
         \label{fig: Artefact_cluster_19_2}
     \end{subfigure}
        \caption{\km{Examples of artefacts that were observed in the results. Large sources (a), bright sources on the edges of images (b) and pre-existing artefacts (c) contributed to the existence of the artefacts observed. The images on the left show the contours that were detected and highlight the resulting crop as a red box. The images in (a) and (b) show the crop that resulted when the detected bounding square crossed the boundary of the image, causing Python to reinterpret the crop according to indexing rules.  The images on the right show the corresponding final images after the preprocessing. The apparent pixelation is caused by resizing small crops that result from attempts to crop out of bounds and missing sources.}}
        \label{fig:Artefact_examples}
\end{figure}

\km{
The types of artefacts detected by the clustering algorithm are a combination of those that existed in the data as well as those introduced by our preprocessing. Cluster 15 contains colour-corrupted images, Cluster 19 contains a mixture of images with bright flares and other artefacts. Clusters 8, 11, 9, 17, 7 and 13 display artefacts introduced when the resizing function fails to locate a galaxy source, leading to a resized image that appears pixelated as shown in \autoref{fig:Artefact_examples}.} 

\km{Comparing the images before and after resizing suggests that this happened in images that either had a pre-existing artefact, bright sources located at the edge of the image or large sources that fill the entire image. This is because our preprocessing assumes that a source of interest is a bright object located at the centre of the image. Our implementation avoids cropping out important galaxy features by detecting a square that is twice the size of the smallest square bounding a contour fit.} 

\km{As our results show, the preprocessing is sufficient for most of the sources in the dataset due to their small sizes relative to the full images. In some cases, however, the detected bounding squares cross the boundaries of an image and may contain negative values in some of their coordinates on an image grid. Attempting to crop such a bounding square produces the results shown in \autoref{fig: Artefact_cluster_13} and \autoref{fig: Artefact_cluster_19}. This is because the crops are obtained by indexing the image as an array, and Python interprets negative indices as counting from the right (or bottom) instead of the left (or top). This results in cropping an image patch that does not contain the galaxy source and is usually much smaller, causing pixelation when the image is resized. Similar pixelation is observed in images that have missing sources, such as \autoref{fig: Artefact_cluster_19_2}. These specific artefacts could have been detected programmatically during preprocessing after they were discovered, but given that most of them were caused by existing artefacts in the image anyway, we elected to leave them in and instead use them as a way to test unsupervised artefact detection. 
}

\section{BYOL Hyperparameter tuning}
\label{appendix:byol_hyperparameters}

\begin{figure}
    \centering
    \includegraphics[width = \linewidth]{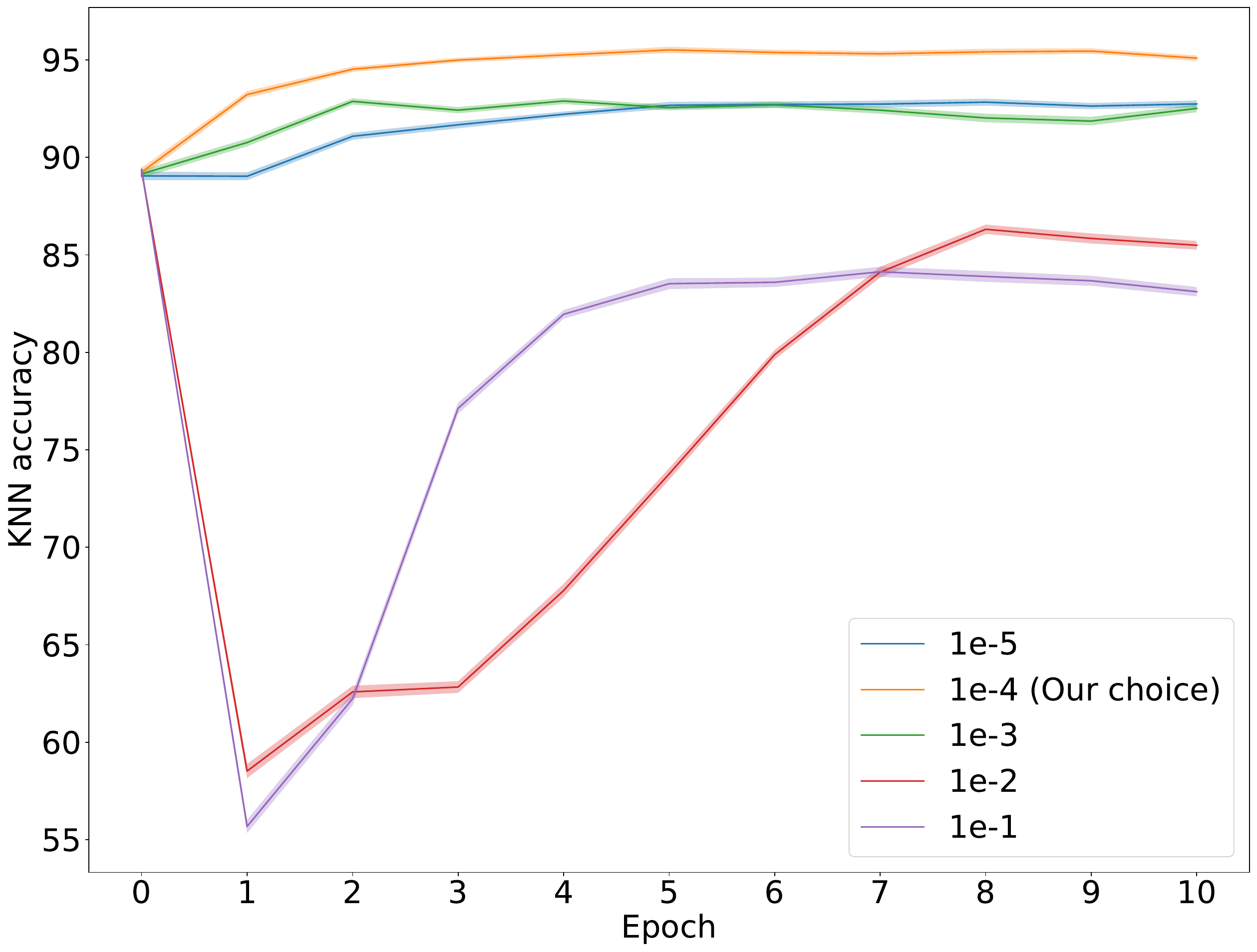}
    \caption{\km{The accuracy of the KNN algorithm for the evaluation subset of the Galaxy Zoo data as a function of epoch for different learning rates.}}
    \label{fig:knn_accuracy_at_different_learning_rate}
\end{figure}

\new{We set most BYOL hyperparameter values to be the same as used in \citet{grill2020bootstrap}, seeing no reason to change them. The batch size differs due to computational restrictions and the number of epochs is selected based on convergence. We did, however, need to change the learning rate in order to obtain rapid convergence to a high quality representation. \autoref{fig:knn_accuracy_at_different_learning_rate} shows the KNN accuracy for the evaluation subset of the Galaxy Zoo data for different learning rates, showing that our choice of $1\times 10^{-4}$ produces excellent performance.}

\new{\autoref{tab:byol_augmentations} shows the effect of using individual augmentations compared with using all of them on the Galaxy Zoo evaluation set. This shows that the exact choice of augmentations did not have a large impact on the accuracy. However, we did find that using all augmentations as described in the text produced the most reliable feature space.}

\begin{table}
    \centering
    \begin{tabular}{cc}
    \hline
    \km{Augmentation} & \km{KNN accuracy} \\
    \hline
    \km{Random resized crops} & \km{95.30 $\pm$ 0.07}\\
    \km{Random rotation} & \km{93.88 $\pm$ 0.07} \\
    \km{Random horizontal and vertical flips} & \km{94.73 $\pm$ 0.07}\\
    \km{Random Gaussian blur} &  \km{94.66 $\pm$ 0.07} \\
    \km{All} &  \km{95.89 $\pm$ 0.07} \\
    \hline
    
    \end{tabular}
    \caption{\km{The effect of individual augmentations on KNN accuracy after 10 epochs.}}
    \label{tab:byol_augmentations}
\end{table}

\section{F1 scores}
\label{appendix:f1}

\begin{table}
    \centering
    \begin{tabular}{ccc}
    \hline
    \km{Class} & \km{F1 score (ImageNet)} & \km{F1 score (Random)}\\
    \hline
    \km{Spirals} & \km{0.9637 $\pm$ 0.0008} & \km{ 0.9531 $\pm$ 0.0009} \\
    \km{Round ellipticals} & \km{0.9526$\pm$ 0.0007} & \km{0.9283 $\pm$ 0.0010}\\
    \km{Edge on galaxies} & \km{0.9464 $\pm$ 0.0008} &  \km{0.9161 $\pm$ 0.0011}\\
    \hline
    \end{tabular}
    \caption{\km{The mean F1 scores and standard errors obtained over 50 seeded runs for the Galaxy Zoo evaluation set, fine-tuning BYOL from ImageNet and Random weights respectively.}}
    \label{tab:f1_galaxy}
\end{table}

\begin{table}

    \centering
    \begin{tabular}{ccc}
    \hline
    \km{Representations} & \km{F1 scores (FRI)} & \km{F1 scores (FRII)}\\
    \hline
    \km{Radio Galaxy Zoo} & \km{0.856 $\pm$ 0.004} & \km{0.870 $\pm$ 0.004}\\
    \km{ImageNet} & \km{0.831 $\pm$ 0.004}  & \km{0.834 $\pm$ 0.005}\\
    \km{BYOL Galaxy Zoo} & \km{0.775 $\pm$ 0.005} & \km{0.792 $\pm$ 0.005}\\
    \km{Random Initial} & \km{0.695 $\pm$ 0.006} & \km{0.722 $\pm$ 0.005}\\

    \hline
    
    \end{tabular}
    \caption{\km{The mean F1 scores and standard errors obtained over 50 seeded runs for the final MiraBest representations.}}
    \label{tab:f1_mirabest}
\end{table}

\new{To further substantiate the performance of our algorithms on the labelled subsets and investigate any class differences that may occur, we make use of the F1 score. The F1 score is the harmonic mean between precision ($p$, the number of true positives divided by the number of predicted positives) and the recall ($r$, the number of true positives divided by the total number of positives in the sample):}

\begin{equation}
\centering
{\rm F1} = 2 \frac{p.r}{p + r}.
\end{equation}
\new{\autoref{tab:f1_galaxy} shows the F1 score per class for the random and ImageNet initialisations of BYOL. This table supports the conclusion that ImageNet weights provide a useful starting point for fine-tuning BYOL. It also shows that this conclusion is not class dependent.}

\new{\autoref{tab:f1_mirabest} shows the F1 scores for the MiraBest dataset for each of the starting weights considered in the text, supporting the conclusions drawn from the accuracy plots. 
}

\section{Cluster analysis}
\new{To provide more numerical evidence for the qualitative results from the clustering analysis presented in \autoref{tab:clusters}, we made use of our evaluation set which had a strict set of cuts applied to the citizen scientist votes to create a very pure subset of three different classes: ellipticals, spirals and edge-on galaxies. It can be see that our qualitative conclusions correspond well to the numerical results from this small subset of data. Almost all the spiral galaxies reside in clusters 0, 5 and 18 while the ellipticals dominate clusters 1, 12 and 14. Clusters 2 and 10 are fairly pure samples of edge-on galaxies while the remaining clusters are more mixed.}
\label{appendix:clusters}
\begin{figure}
    \centering
    \includegraphics[width = 0.8\linewidth]{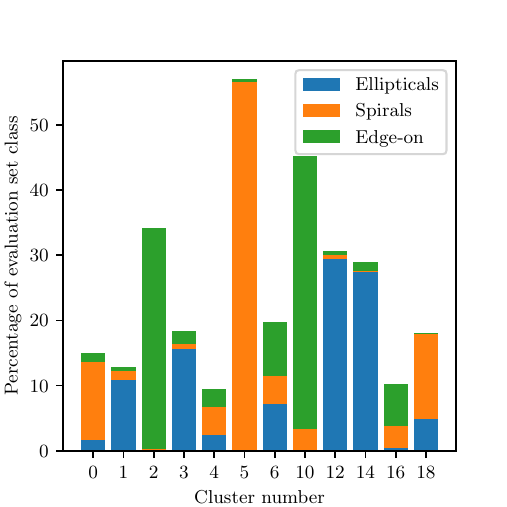}
    \caption{\new{The break down of what percentage of each class in the evaluation set lies in which cluster. The quantitative results using the ``clean'' evaluation set broadly agree with the qualitative conclusions from visual inspection displayed in \autoref{tab:clusters} and the broader vote distributions in \autoref{fig:distributions}. Artefact clusters are excluded.}}
    \label{fig:evaluation_clusters}
\end{figure}

\bsp	
\label{lastpage}
\end{document}